\documentclass[aps, prd, amsmath, floats, floatfix, reprint,superscriptaddress, nofootinbib, showpacs,10pt]{revtex4-1}

\usepackage{graphicx}

\usepackage{amsfonts}
\usepackage{subfigure}
\usepackage{xspace} 
\usepackage[usenames,dvipsnames]{color}
\usepackage{dcolumn}
\usepackage{bm}
\usepackage{hyperref}
\usepackage{mathrsfs}
\usepackage[]{amsmath,amssymb}
\usepackage[]{amsthm}
\usepackage{psfrag}
\usepackage{comment}

\theoremstyle{plain}

\theoremstyle{definition}

\newcommand{\norm}[1]{\lVert#1\rVert}

\usepackage{ulem}
\newcommand{\Cornell}{\affiliation{Center for Radiophysics and Space
    Research, Cornell University, Ithaca, New York, 14853}}
\newcommand{\CITA}{\affiliation{Canadian Institute for Theoretical Astrophysics,
    University~of~Toronto, Toronto, Ontario M5S 3H8, Canada}}
\newcommand{\CIFAR}{\affiliation{Fellow, Canadian Institute for Advanced Research}}
\newcommand{\DAA}{\affiliation{Department of Astronomy and Astrophysics, University of Toronto, Toronto, Ontario M5P 3H4, Canada}}
\definecolor{darkgreen}{rgb}{0.2,0.7,0.2}

\newcommand{\w}[1]{\bm{#1}}
\begin{document}

\title[Precession - tracking coordinates for simulations of
compact-object-binaries.]{Precession--tracking coordinates for
  simulations of compact--object--binaries}

\author{Serguei Ossokine}\CITA\DAA
\author{Lawrence E. Kidder} \Cornell
\author{Harald P. Pfeiffer} \CITA\CIFAR

\begin{abstract}
  Binary black hole simulations with black hole excision using
  spectral methods require a coordinate transformation into a
  co-rotating coordinate system where the black holes are essentially
  at rest. This paper presents and discusses two coordinate
  transformations that are applicable to {\em precessing} binary
  systems, one based on Euler angles, the other on quaternions. Both
  approaches are found to work well for binaries with moderate
  precession, i.e. for cases where the orientation of the orbital
  plane changes by $\ll 90^\circ$. For strong precession, performance
  of the Euler-angle parameterization deteriorates, eventually failing
  for a $90^\circ$ change in orientation because of singularities in
  the parameterization (``gimbal lock''). In contrast, the quaternion representation is
  invariant under an overall rotation, and handles any orientation of
  the orbital plane as well as the Euler-angle technique handles
  non-precessing binaries. 
  
\end{abstract}

\date{\today}


\maketitle

\section{Introduction}
\label{sec:intro}

Gravitational waves offer an exciting new observational window into
the universe. With the second generation of gravitational wave
detectors such as Advanced LIGO and Advanced Virgo commencing
observations in 2015 \cite{Aasi:2013wya}, it is extremely important to
develop a detailed picture of the gravitational physics of the most
likely sources. A very promising source of gravitational waves are
inspiraling and merging binary black holes~\cite{Abadie:2010cfa}.
Because of the weakness of the gravitational wave signal, matched
filtering is necessary to pick out the waveform from the noise
~\cite{Finn1992,Finn1993}. Constructing such templates, in turn,
requires direct numerical integration of Einstein's equations for the
late inspiral, merger and ringdown phase of the coalescing compact
object binary; see, e.g.~\cite{Ohme:2011rm}. Since 2005, starting with
the seminal work of Frans Pretorius \cite{Pretorius2005a}, many groups
have successfully simulated binary black hole systems using a variety
of different techniques. For recent overviews of the state of the
field, see \cite{Hinder:2010vn, Pfeiffer:2012pc}.

Compact object inspirals fall into two categories: non-precessing and
precessing. While the non-precessing aligned-spin systems arguably
represent an important subspace of all binary black hole systems, the
more general case features arbitrary spin orientations. In this
non-symmetric situation, the interaction of the orbital angular momentum and
the black holes' spins leads to precession of the orbital plane,
changing its orientation by as much as 180 degrees.

Precession modulates the gravitational waveform. Therefore, it is
crucial to explore these strongly precessing systems. Furthermore,
precessing systems allow the study of gravitational dynamics in an 
underexplored regime, providing a new opportunity for comparing
numerical relativity to various analytic approximations like
post-Newtonian (see e.g. \cite{lrr-2002-3,Blanchet:2009rw}) and
effective-one-body theory (see e.g.
\cite{Buonanno99,Damour01c,DIN,Damour:2012mv}). The numerical
simulations can be used both to test the accuracy of the analytic
treatments and to calibrate them, in some cases, thus improving their
accuracy~\cite{Damour2009a,Buonanno:2009qa,Taracchini:2012}.
Furthermore, one can attempt to reproduce numerically predictions from
analytic computations such as transitional
precession~\cite{Apostolatos1994}, which is known from PN theory but
has not yet been observed in numerical simulations.

Numerical simulations of precessing binary black holes have already
been undertaken; for example
\cite{Campanelli2007b,CampanelliEtal2009,Sturani:2010ju,Sturani:2010yv,Schmidt2010,Zlochower:2010sn,Lousto:2011kp,Lousto:2012gt}.
Given the vastness of parameter space and the need for simulations
lasting at least 10 orbits - possibly 100's of orbits - to optimally
exploit gravitational wave
detectors~\cite{Ohme:2011rm,Damour:2010,Santamaria:2010yb,Boyle:2011dy,OhmeEtAl:2011,MacDonald:2012mp},
a lot of extra work remains to be done.

The Spectral Einstein Code {\tt SpEC}~\cite{SpECwebsite} is allows
efficient and accurate simulations of binary black holes; see
e.g.~\cite{Boyle2007,Scheel2009,Chu2009,Buchman:2012dw,Lovelace:2011nu,Lovelace:2010ne,MacDonald:2012mp,Mroue:2012kv}.
This code applies black hole excision and uses time-dependent
coordinate mappings to rotate and deform the computational grid such
that the excision regions remain inside the black hole horizons at all
times. For non-precessing inspiralling binaries, these coordinate
mappings are described in detail in previous
work~\cite{Boyle2007,Buchman:2012dw,Hemberger:2012jz}.  

The purpose of the present paper is to develop coordinate mappings
that are able to follow a precessing compact object binary through the
inspiral, even for strongly precessing systems. We present two
different approaches. The first one is based on Euler angles; it works
well for moderate precession, but fails when the orientation of the
orbital plane changes by 90 degrees or more. The second approach is
designed to avoid the deficiencies of the Euler angle
parameterization. By using quaternions, we devise coordinate mappings
that work for any change of orientation of the orbital plane with a
performance comparable to the earlier non-precessing techniques. The
techniques developed here have already been used
in~\cite{PhysRevD.87.084006,Mroue:ManyMergersPRL}

This paper is organized as follows. Section~\ref{sec:methods}
describes the computational setup of {\tt SpEC} in more detail
(Sec.~\ref{sec:DualFrames}), and develops the coordinate mappings
based on Euler angles (Sec.~\ref{sec:genDeriv}) and quaternions
(Sec.~\ref{sec:QuaternionControl}). Section~\ref{sec:Results} presents
a sequence of numerical results obtained with both approaches,
starting from Newtonian and post-Newtonian test-cases to simulations
of binary black holes with numerical relativity (NR). We summarize our
results in Sec.~\ref{sec:Discussion}.

\section{Methods and Techniques}
\label{sec:methods}

\subsection{Dual frames and control systems}
\label{sec:DualFrames}

As described in Scheel et al.~\cite{Scheel2006}, {\tt SpEC} 
utilizes a dual-frame approach to simulate compact object binaries.
Einstein's equations are written down in an asymptotically
non-rotating coordinate-system $x^{\bar a}=(\bar t, x^{\bar\imath})$,
referred to as the ``inertial frame'', and all tensors are represented
in the coordinate basis of this frame.  In the inertial frame, tensor
components remain finite even at large separation.  The computational
grid is specified in ``grid coordinates'' $x^a=(t,x^i)$.  The
collocation points of the spectral expansion are at constant grid
coordinates, and numerical derivatives are computed with respect to
these coordinates.  The two coordinate frames share the same
time-coordinate
\begin{equation}
\bar t=t.
\end{equation}
The spatial coordinates of the two frames are related by a coordinate
transformation
\begin{equation}\label{eq:CoordTrafo}
  x^{\bar\imath}=x^{\bar \imath}\left(x^i; \lambda^\mu(t)\right),
\end{equation}
which depends on a set of parameters $\lambda^\mu(t)$ to be discussed
in detail later.  The coordinate
transformation Eq.~(\ref{eq:CoordTrafo}) maps the grid-coordinates
into the inertial frame such that the excision surfaces (coordinate
spheres in the grid-frame) are mapped to locations somewhat inside
the apparent horizons of the black holes in the inertial frame.

In the original work~\cite{Scheel2006}, this coordinate transformation
was taken as the composition of a rotation about the z-axis and an
overall scaling of the coordinates\footnote{Note that in Ref.~\cite{Scheel2006},
the equations give the transformation from the inertial coordinates to
the grid coordinates, the rotation angle is $\phi$ instead of $\psi$,
and the scale factor $a$ is the inverse of the scale factor in this
paper.},
\begin{equation}\label{eq:xyRotation}
x^{\bar\imath}=
\left(
\begin{aligned}
\bar x \\ \bar y \\ \bar z
\end{aligned}
\right)
=a(t)\left(
\begin{aligned}
\cos{\psi(t)} && - \sin{\psi(t)} && 0 \\
\sin{\psi(t)} && \cos{\psi(t)}&& 0 \\
0& & 0 && 1
\end{aligned}
 \right)
\left(
\begin{aligned}
x \\ y \\ z
\end{aligned}
\right).
\end{equation}

In this simple case, the map $\lambda^\mu(t)=\{a(t), \psi(t)\}$
depends on two parameters: the scale factor $a(t)$ and the rotation
angle $\psi(t)$. The map parameters $\lambda^\mu(t)$ are chosen
dynamically during the simulation, such that the map tracks the actual
motion of the black holes. This can be accomplished by introducing a
set of control-parameters $Q^\mu$, such that
\begin{enumerate}
\item $Q^\mu=0$ if the mapped excision spheres are at the desired
  location in inertial coordinates.
\item Under small variations of the mapping parameters around their current values, the control-errors satisfy
\begin{equation}\label{eq:Orthogonality}
\left.\frac{\partial Q^\mu}{\partial \lambda^\nu}\right|
_{\lambda^\mu=\lambda^\mu(t)}
= -\delta^\mu_\nu
\end{equation}
\end{enumerate}
While not strictly required, Eq.~(\ref{eq:Orthogonality}) allows one
to write down uncoupled feedback control equations for the
$\lambda^\mu(t)$. In the special case of a linear, uncoupled system,
this reduces to $Q^\mu = \lambda^\mu_{\mathrm target} - \lambda^\mu$.

For black holes orbiting in the xy-plane, Eq.~(\ref{eq:xyRotation})
suffices to keep the excision boundaries inside the inspiraling black
holes, resulting in successful simulations of inspiraling BH--BH
binaries in Ref.~\cite{Pfeiffer-Brown-etal:2007}. Subsequently, the
map was refined to avoid a rapid inward motion of the outer
boundary~\cite{Scheel2009}, to adjust the shapes of the mapped
excision boundaries to more closely conform to the distorted apparent
horizons~\cite{Szilagyi:2009qz,Lovelace:2011nu,Lovelace:2010ne}, and
was generalized to unequal mass binaries~\cite{Buchman:2012dw}.
Hemberger et al~\cite{Hemberger:2012jz} summarizes these maps, and
introduces further mappings that are needed during the merger phase of
the black hole binary.

The purpose of the present paper is the development of coordinate
mappings that can handle precessing binaries.  Because in general the
center of mass will move (e.g due to asymmetric GW emission), these
coordinate mappings must also allow for a translation of the binary.
Rotation and translation couple to each other and must therefore be
dealt with simultaneously. The questions addressed in this paper are
therefore (1) determination of a suitable coordinate mapping for
precessing, translating binaries, (2) suitable choice of mapping
parameters $\lambda^\mu$, and (3) derivation of control-parameters
$Q^\mu$. Specifically, we will discuss below two generalizations of
Eq.~(\ref{eq:xyRotation}), one based on Euler-angles and one based on
quaternions. We will show that the Euler-angle representation suffers
from singularities when the inclination of the orbital plane passes
through $\pi/2$, and we will demonstrate that the quaternion
representation fixes these problems.

\subsection{Euler angle representation} 
\label{sec:genDeriv}

In the general case where the orbital plane precesses, we use a
mapping that composes a scaling $a(t)$, a rotation $R(t)$ and a
translation $\vec{T}(t)$.  The mapping is given by
\begin{equation}\label{eq:3dMap}
\vec{\bar{x}} = a(t)\, R(t)\, \vec{x} + \vec{T}.
\end{equation}
A rotation matrix can be specified by Euler angles,
\begin{widetext}
\begin{equation}\label{eq:EulerRotation}
R = \begin{pmatrix}
\cos{\theta} \cos{\psi} & 
- \cos{\phi} \sin{\psi} + \sin{\phi} \sin{\theta} \cos{\psi} &
\sin{\phi} \sin{\psi} + \cos{\phi} \sin{\theta} \cos{\psi} \\
\cos{\theta} \sin{\psi} & 
\cos{\phi} \cos{\psi} + \sin{\phi} \sin{\theta} \sin{\psi} & 
- \sin{\phi} \cos{\psi} +\cos{\phi} \sin{\theta} \sin{\psi}  \\
-\sin{\theta} &
\sin{\phi} \cos{\theta} &
\cos{\phi} \cos{\theta} 
\end{pmatrix}.
\end{equation}
\end{widetext}
where $\phi$ is the roll angle around the $x$-axs, $\theta$ is the
pitch angle around the $y$-axis, and $\psi$ is the yaw angle around the
$z$ axis and we have suppressed the explicit time-dependence.  

For our application, the desired locations of the black holes lie
parallel to the $x$-axis, i.e. the black holes are at grid coordinates
$(c^x_A,c^y,c^z)$ and $(c^x_B,c^y,c^z)$.  It is straightforward to
show that for these two points a rotation about the $x$-axis is
degenerate with a translation {\em because only the location of the
  black holes is important}.  Therefore, we can set $\phi(t)=0$ so
that\footnote{Note that if the motion is confined to the $x-y$ plane,
  the pitch will remain fixed at $\theta = 0$ and we recover the
  rotation matrix in Eq.~(\ref{eq:xyRotation}).}
\begin{equation}\label{eq:PitchYawRotation} R = \left(
\begin{aligned}
\cos{\theta} \cos{\psi} && - \sin{\psi} && \sin{\theta}
\cos{\psi} \\ \cos{\theta} \sin{\psi} && \cos{\psi} &&
\sin{\theta} \sin{\psi} \\ -\sin{\theta} & & 0 &&
\cos{\theta}
\end{aligned}
 \right).
\end{equation}
Thus the mapping in Eq.~(\ref{eq:3dMap}) will have six parameters in
this case, a scaling $a(t)$, a pitch angle (rotation about y-axis)
$\theta(t)$, a yaw angle (rotation about z-axis) $\psi(t)$, and a
translation $(T^X(t),T^Y(t),T^Z(t))$.

The goal of the scaling-rotation-translation map is to keep the
horizons of the two black holes centered on the excision surfaces. As
the binary evolves, the map parameters need to be adjusted by the
control system. To derive the control parameters $Q^{\mu}$, c.f.
Eq(\ref{eq:Orthogonality}), one can consider perturbations of the
mapping parameters around their current values. Let $\lambda^\mu =
\{a,\theta,\psi,\vec{T}=(T^X,T^Y,T^Z)\}$ be the current imperfect
mapping parameters at some time during the evolution. Furthermore,
denote the desired parameters
$\lambda_{0}^{\mu}=\lambda^{\mu}+\delta\lambda^{\mu}=\{a_{0},\theta_0,\psi_{0},T_{0}^{X},T_{0}^{Y},T_{0}^{Z}\}$.
Finally, let $\vec{x}_{A}$ and $\vec{x}_B$ denote the current location
of the center of black hole A and B, respectively\footnote{The precise
  definition of ``center'' is not important; we shall use the
  coordinate point around which the coordinate radius of the apparent
  horizon has vanishing $l=1$ multipoles.}, and let $\vec c_A$ and
$\vec c_B$ denote the desired location of the black hole centers; i.e.,
the centers of the excision spheres. For convenience we also define
the vectors $\vec{X} = \vec{x}_{A}-\vec{x}_{B}$ and
$\vec{C}=\vec{c}_{A}-\vec{c}_{A}$. The target mapping
$\lambda_{0}^{\mu}$ is such that the points $\vec{c}_{A},\
\vec{c}_{B}$ are mapped onto the inertial frame position of the black
holes,
$\vec{\bar{c}}_{A}=\vec{\bar{x}}_{A},\;\vec{\bar{c}}_{B}=\vec{\bar{x}}_{B}
$:
\begin{equation}
{\vec{\bar{x}}}_{\mathrm A,B} = a_{0}
R({\theta_{0}},{\psi_{0}}) {\vec{c}}_{\mathrm A,B} + \vec{T_{0}}.
\end{equation}
Rewriting this equation in terms of the current mapping and the grid
location of each black hole (i.e $\vec{x}_{A,B}$) yields
\begin{equation}
\label{eq:generalSys}
a R(\theta,\psi) {\vec{x}}_i + \vec{T} = (a + \delta a)
R(\theta + \delta\theta,\psi+\delta\psi) {\vec{c}}_{i} +
\vec{T} + \vec{\delta T}.
\end{equation}
where $i=A,\ B$.

Equation (\ref{eq:generalSys}) represents six equations for the six
unknowns $\delta\lambda^{\mu}$. Solving this system of equations to
leading order in the perturbations yields
\begin{subequations}
\begin{eqnarray}
\delta a &=& a \left( \frac{X^{x}}{C^{x}} - 1 \right), \label{eq:Qa} \\
\delta \theta &=& \frac{-X^{z}}{C^{x}}, \label{eq:QPitch}\\
\delta \psi &=& \frac{1}{\cos{\theta}} \frac{X^{y}}{X^{x}}, \label{eq:QYaw} \\
\delta T^X &=& \frac{a}{C^{x}} \left( 
   \delta t^X \cos{\theta} \cos{\psi} - \delta t^Y \sin{\psi} \right. \label{eq:QTx}
\nonumber \\ && + \left. 
   \delta t^Z \sin{\theta} \cos{\psi} \right),\\
\delta T^Y &=& \frac{a}{C^{x}} \left( 
   \delta t^X \cos{\theta} \sin{\psi} + \delta t^Y \cos{\psi} \right. \label{eq:QTy} 
\nonumber \\ && + \left. 
   \delta t^Z \sin{\theta} \sin{\psi} \right),\\
\delta T^Z &=& \frac{a}{C^{x}} \left( - \delta t^X \sin{\theta} +
                                              \delta t^Z \cos{\theta}
                                            \right) \label{eq:QTz}
\end{eqnarray}
\end{subequations}
where
\begin{subequations}
\begin{eqnarray}
\delta t^X &=& c^x_A x_B - c^x_B x_A + c^y  X^{y}  + 
               c^z X^{z}, \nonumber \\ && \\
\delta t^Y &=& c^x_A y_B - c^x_B y_A - c^y X^{x} \nonumber 
\\ && - ~ c^zX^{y} \tan{\theta},\\
\delta t^Z &=& c^x_A z_B - c^x_B z_A + c^y X^{y}
               \tan{\theta} \nonumber \\ && + ~ c^z X^{x},
\end{eqnarray}
\end{subequations}
Furthermore, we have assumed that the centers of the excision surfaces
are aligned parallel to the $x$-axis so that $c_A^y = c_B^y = c^y$ and
$c_A^z = c_B^z = c^z$. 

Perhaps surprisingly, the $\delta\lambda^{\mu}$ given by
Eq.~(\ref{eq:Qa} --\ref{eq:QTz}) are the desired control parameters
$Q^{\mu}$. This can be seen as follows. For a perfect map,
$\lambda^{\mu}=\lambda_{0}^{\mu}$, i.e.
$\delta\lambda^{\mu}=Q^{\mu}=0$. Moreover, by definition

\begin{equation}
  \frac{\partial Q^{\mu}}{\partial\lambda^{\nu}}=\frac{\partial}{\partial \lambda^{\nu}}\left(\lambda_{0}^{\mu}-\lambda^{\mu}\right)=\frac{\partial \lambda_{0}^{\mu}}{\partial \lambda^{\mu}} -\frac{\partial \lambda^\mu}{\partial \lambda^\nu }=-\delta^{\mu}_{\nu}.
\end{equation}
Thus $\delta \lambda^{\mu}$ defined by Eq.~(\ref{eq:Qa} --\ref{eq:QTz})
satisfy the conditions for $Q^\mu$ outlined in section
\ref{sec:DualFrames}.

\begin{figure}
  \includegraphics[width=\columnwidth]{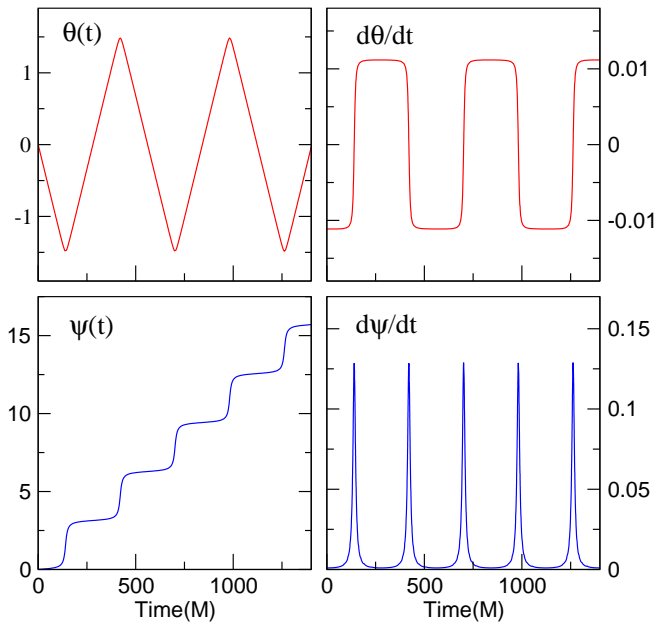}
  \caption{Typical behaviour of the Euler angles and their derivatives for a nearly polar orbit inclined at 85 degrees with respect to the $x-y$ plane }
  \label{fig:PitchYawBroken}
\end{figure}

The Euler angle prescription as described above is adequate for
describing rotations that are close to the $x-y$ plane, and has been
used for the {\tt SpEC} simulations presented
in~\cite{Szilagyi:2009qz}. However, the Euler angle prescription
carries with it an inherent coordinate singularity that causes a
breakdown of the control system for high inclination angles. Firstly,
note that Eq.~(\ref{eq:QYaw}) shows that $\delta \psi$ will diverge
when $\theta=\frac{\pi}{2}$, which would lead to the breakdown of a
feedback control system. Further, since {\tt SpEC} uses
proportional-derivative-control or
proportional-integral-derivative-control~\cite{Hemberger:2012jz}, we
must examine the behaviour of the derivatives of the Euler angles,
$\dot{\theta},\ \dot{\psi}$. Notice that we can relate the angular
velocity to the derivatives of the Euler angles simply by the
relationship
\begin{equation}
  \label{eq:velsys}
(\omega_1,\omega_2,\omega_3)^{T}=A\ (\dot{\phi},\dot{\theta},\dot{\psi})^{T},
\end{equation}
where $A$ is the Euler angle rates matrix:

\begin{equation}
A=\begin{pmatrix}
\cos\psi\cos\theta & -\sin\psi & 0 \\
\sin\psi\cos\theta & \cos\psi & 0 \\
-\sin\theta & 0 & 1

\end{pmatrix}.
\end{equation}
The black holes move on regular trajectories with a slowly varying
orbital frequency $\vec\omega$ in inertial coordinates; therefore, the
left-hand side of Eq.~(\ref{eq:velsys}) is continuous. However,
$|\det{A}|=|\cos(\theta)|$, so that for $\theta=\frac{\pi}{2}$ the
time-derivatives of the Euler angles will diverge since, by Cramer's
rule, the inverse of $A$ scales as $1/\det{A}$.

There is another way to envision the divergence of the derivatives of
the Euler angles. Consider the unit vector in the direction connecting
the centers of the two compact objects in inertial coordinates,
$\hat{u}=\frac{\vec{\bar{x}}_B-\vec{\bar{x}}_A}{|\vec{\bar{x}}_B-\vec{\bar{x}}_A|}$. Before,
we considered $\psi,\ \theta$ as parameters in a mapping. Let us now
consider them as spherical polar coordinates that describe this
vector\footnote{$\theta$ here means the angle to the $xy$ plane rather
then the angle to the $z$-axis. Therefore the following equations
differ slightly from standard spherical polar coordinates.}
\begin{eqnarray}
\left(
\begin{aligned}
u_x \\  u_y \\ u_z
\end{aligned}
\right) =\left(\begin{aligned}
 \cos\psi\cos\theta\\
\sin\psi\cos\theta\\
\sin\theta\end{aligned}\right).
\end{eqnarray}
We can immediately derive the expressions for $\dot{\theta},\
\dot{\psi}$ as functions of $\dot{u}_x, \ \dot{u}_y,\
\dot{u}_z$:
\begin{eqnarray}
  \dot{\theta} &=& \frac{\dot{u}_{z}}{\cos\theta},\\
  \dot{\psi}   &=& \frac{1}{\cos\theta}\sqrt{\dot{u}_{x}^{2}+\dot{u}_{y}^2-\dot{u}_{z}^2\tan^{2}\theta}.
\end{eqnarray}
From these equations it is obvious that the derivatives of $\theta$
and $\phi$ behave abnormally when $\hat u$ moves across one of the poles at
uniform velocity. In fact, $\dot{\theta}$ does not exist, and its
second derivative diverges (see the top panels of Fig.~\ref{fig:PitchYawBroken}). Meanwhile, $\dot{\psi}$ diverges: letting
$\delta\theta = \frac{\pi}{2}-\theta$ we can write, for $\delta\theta\ll 1$:

\begin{equation}
\dot{\psi} \propto
\frac{\omega}{\delta \theta}.
\end{equation}

This behaviour is demonstrated clearly in Figure
\ref{fig:PitchYawBroken}, where the derivatives of both Euler angles
angles demonstrate sharp and nearly discontinuous features.\\

This is the fundamental reason why Euler angles are not a suitable
parametrization of rotations: there exist situations, which we would
like to study, when their derivatives grow extremely fast numerically.

\subsection{Rotation--invariant Quaternion representation}
\label{sec:QuaternionControl}

The origin of the break-down of the Euler angle
  representation lies in its reliance of a preferred coordinate system,
  which is implicit in the adoption of Euler angles.  The physics of
  compact object inspirals is invariant under rotations of the spatial
  coordinates.  Ideally, the numerical methods used to describe such a
  system should also be invariant, and should work equally well
  independent of the orbital plane of the black holes.  

  The singularities in the Euler angle representation arise from a
  poor choice of representation of the rotation group, which relied on
  preferred directions in space (namely the coordinate axes).
  Therefore, a suitable representation must be independent of any
  special directions. We employ quaternions to represent rotations and
  build up the overall rotation from a sequence of infinitesimal
  rotations.

It should be noted that a similar construction can be done with a
different paramtetrization of rotations. See the Appendix for an
example using orthogonal infinitesimal rotation matrices.

\subsubsection{Quaternion algebra}
\label{sec:quatAlg}

Quaternions are an extension of the complex numbers, with three
imaginary units $i, j$, and $k$, obeying
\begin{equation}
i^{2}=j^{2}=k^{2}=ijk=-1,
\end{equation}
as well as certain further multiplication rules. A quaternion $\w{q}$
has the form
\begin{equation}
\w{q}=q_{0}+q_{1}i+q_{2}j+q_{3}k,\quad\ q_0,\ldots,q_3 \in \mathbb{R}
\end{equation}

This is conveniently written as $\w{q} = (q_{0},\vec{q}) $, where
$\vec q=(q_1,q_2,q_3)$. Addition and scalar multiplication are defined
in analogy with complex numbers. With the structure introduced so far,
the set of quaternions
\begin{equation}
\mathbb{H}=\{q_0 + q_1 i +q_2 j +q_3 k | q_{i}\in \mathbb{R} \} 
\end{equation}
is a 4-dimensional vector space over the real numbers. Multiplication
is defined by
\begin{equation}
\w{qp} = (p_{0}q_{0}- \vec{p}\cdot\vec{q},
p_{0} \vec{q} + q_{0} \vec{p} + \vec{p} \times \vec{q}),
\end{equation}
where $\vec{q}\cdot\vec{q}$ and $\vec{q}\times\vec{p}$ are the
standard Euclidean dot and cross products respectively.
Complex conjugation is given by
\begin{equation}
\w{q}^{*}=(q_{0},-\vec q).
\end{equation}
It follows that the multiplicative inverse is given by 
\begin{equation}
\w{q}^{-1}=\frac{\w{q}^{*}}{|\w{q}|}, 
\end{equation}
where the norm $|\w{q}|$ satisfies
\begin{equation}
  |\w{q}|^2 = \w{q}\,\w{q}^*= q_0^2 + \vec{q}{\,}^2.
\end{equation}

Restricting our attention now to the set of all unit quaternions $
Sp(1) = \{\w{q}\in \mathbb{H},|\w{q}|=1\} $, it is easy to show that
$Sp(1)$ is isomorphic to $SU(2) $ where $ SU(2) $ is the group of all
$ 2 \times 2 $ unitary matrices with unit
determinant~\cite{altmann2005rotations}. $SU(2)$ is a double cover of
the rotation group $SO(3)$, which means that unit quaternions do
represent rotations.

Unit quaternions are related to rotations in the following manner. Let
$\hat{n}$ be a unit-vector, and define
\begin{equation}
\label{eq:rotQuat}
\w{q}=(\cos\frac{\theta}{2},\hat{n}\sin\frac{\theta}{2})
\end{equation}
for some angle $\theta$. The quaternion $\w{q}$ rotates a vector
$\vec{v}$ into the vector $\vec{v}'$, around the axis $\hat{n}$ by
angle $\theta$ in the right-handed sense via
\begin{equation}\label{eq:Quaternion-Rotation}
\w{v}' = \w{qvq}^*.
\end{equation}
In this equation, 3-vectors are to be promoted to quaternions by the
rule $\w{v}=(0,\vec{v})$, $\w{v}'=(0,\vec v\,')$,and $|\w{q}|=1$ implies
that $\w{v}'$ has indeed a vanishing real part.
Equation~(\ref{eq:Quaternion-Rotation}) is equivalent to
$\vec{v}\,'=R_{\w{q}}\vec{v}$, with rotation matrix
\begin{equation}
\!\!\!\!\!\!\!R_{\w{q}}\!=\!\!\left(\begin{aligned}
q_0^2\!+\!2q_1^2\!-\vec{q}\,{}^2 &\!& 2(q_1q_2\!-\!q_0q_3) &\!& 2(q_0q_2\!+\!q_1q_3)\\
2(q_1q_2\!+\!q_0q_3) &\!& q_0^2\!+\!2q_2^2\!-\vec{q}\,{}^2 &\!& 2(q_2q_3\!-\!q_0q_1)\\
2(q_1q_3\!-\!q_0q_2) &\!& 2(q_0q_1\!+\!q_2q_3) &\!&  q_0^2\!+\!2q_3^2\!-\vec {q}\,{}^2\\
\end{aligned}\right)
\end{equation}

We can now rewrite Eq.~(\ref{eq:3dMap}) in quaternion language, 
\begin{equation}
  \label{eq:SpECMap}
  \w{\bar{x}}=a\,\w{qxq^* + T}.
\end{equation}
Here, $\vec x,\vec{\bar{x}},\vec T$ have been promoted to quaternions; e.g., $\w{T}=(0,\vec T)$.

Our next task is to derive equations that determine the time evolution
of $\w{q}$ as well as the control parameters $Q^{\mu}$.

\subsubsection{Quaternion kinematics}
\label{sec:quatkin}

In this section we derive the differential equation obeyed by the
rotation quaternion $\w{q}$. Consider a time-dependent unit-quaternion
$\w{q}(t):\mathbb{R} \rightarrow Sp(1)$. The derivative is defined by
\begin{equation}
\dot{\w{q}} = \lim_{h\rightarrow 0}\frac{\w{q}(t+h)-\w{q}(t)}{h}
\end{equation}
We write the rotation  $\w{q}(t+h)$ at time $t+h$, as a product of
$\w{q}(t)$ and a quaternion $\w{u}$ representing an
infinitesimal rotation, 
\begin{equation}
  \w{q}(t+h)=\w{u}\,\w{q}(t).
\end{equation}  The quaternion $\w{u}$ is easily obtained by expanding
the right hand side of Eq.~(\ref{eq:rotQuat}) to first order in
$\theta$ using $\cos\frac{\theta}{2} \approx 1,
\ \sin\frac{\theta}{2} \approx \frac{\theta}{2}$:
\begin{equation}\label{eq:InfinitesimalRotation}
\w{u} = (1,\hat{n}\frac{\theta}{2})=1+\w{\delta q}/2 
\end{equation}
with $\w{\delta q}=(0,\hat{n}\theta)$. If the rotational velocity is
$\vec\omega$ in inertial coordinates, then $\hat n=\hat\omega$ and
$\delta\theta=\omega\,h$, so that $\w{\delta q}=\w{\omega}\,h$. Thus we
can write $\w{q}(t+h)=(I+\w{\delta q}/2) \w{q}(t)$. Substituting,
\begin{equation}
\dot{\w{q}} 
= \lim_{h\rightarrow 0}\frac{(I+\w{\delta q}/2)\w{q}(t) - \w{q}(t)}{h}
= \frac{1}{2}\w{\omega}\,\w{q}.
\end{equation}

Noting that in grid coordinates the angular velocity is $\w{\Omega} =
\w{q}^{*}\w{\omega} \w{q}$, we finally obtain
\cite{Arribas2006}\footnote{This reference uses the opposite
  convention of the one adapted here: $\Omega$ is the angular velocity
  in the fixed frame whereas $\omega$ is the angular velocity in the
  rotating frame.}
\begin{equation}
\label{eq:qderiv}
\dot{\w{q}} = \frac{1}{2} \w{q\Omega}.
\end{equation}

\subsubsection{Quaternion control system}
\label{sec:constr}

The goal of the sector of the control-system for rotations is to keep
the vector $\vec{X}$ \ parallel to the vector $\vec{C}$.
The misalignment between them can be measured by the
  rotation needed to make these vectors parallel:
\begin{equation}
\label{eq:naiveResult}
\vec
Q_R=\frac{\vec{C}\times\vec{X}}{\norm{\vec{C}}^{2}},
\end{equation}

where the subscript `R' indicates that this quantity is of relevance
for rotations.\footnote{The normalization chosen corresponds to the
  fact that only the direction of the two vectors matter in the
  context of rotations, and due to the scaling control system we
  should have to first order, $\norm{\vec{X}}\simeq\norm{\vec{C}}$.}
The control-system needs to adjust the angular velocity $\vec\Omega$
such that $\vec{Q}_R\approx 0$. As long as the control-system works,
this instantaneous rotation is small, and therefore, non-commutativity
of rotations can be neglected. This suggests to control the angular
velocity in the moving frame $\vec\Omega$ based on the
control-parameter $\vec{Q}_R$.

We proceed as follows: We measure $\vec{Q}_R$ regularly during the BBH
evolution,and compute its first and second time-derivatives. As in
earlier work~\cite{Scheel2006} (and in many papers
since~\cite{Lovelace:2010ne,Chu2009,Boyle2007,Pfeiffer-Brown-etal:2007,Scheel2009,Buchman:2012dw,Mroue:2012kv}),
we use this to reset the third time-derivative of the
mapping-parameters that determine the rotation. These parameters are
the second time-derivative of $\vec\Omega(t)$; thus, we choose
$\vec\Omega(t)$ such that it has constant second time-derivative. We
periodically reset this constant using the equation

\begin{equation}\label{eq:d2Omega-reset}
  \frac{d^2\vec\Omega}{dt^2}=\alpha \vec Q_R + \beta \frac{d\vec Q_R}{dt} + \gamma\frac{d^2\vec{Q}_R}{dt^2}.
\end{equation}
A constant value of $d^{2}\vec\Omega/dt^2$ implies that $\Omega(t)$ is
a piece-wise quadratic polynomial. Whenever the second derivative is
reset, we choose integration constants such that $\vec\Omega$ and
$d\vec\Omega/dt$ are continuous. Finally, we use $\vec{\Omega}(t)$ to
determine the actual rotation-matrix via Eq.~(\ref{eq:qderiv}).

There are alternative control-feedback equations to
Eq.~(\ref{eq:d2Omega-reset}). Some of them are discussed
in~\cite{Hemberger:2012jz}. The details of the feedback equation do
not influence the main focus of this paper which is how to represent
rotations and control parameters.

In {\tt SpEC}, Eq.~(\ref{eq:qderiv}) is integrated with a 5-th order
Dormand-Prince time-stepper~\cite{Press2007}.

While Eq.~(\ref{eq:qderiv}) analytically preserves the unit-norm of
$\w{q}$, numerical integration will not identically preserve
$|\w{q}|=1$. Therefore, the $\w{q}(t)$ returned by the ODE-integrator
is rescaled to unit-length, $\w{q}\to \w{q}/|\w{q}|$ before it is used
to construct rotations.

Whenever $d^2\vec\Omega/dt^2$ is reset via
Eq.~(\ref{eq:d2Omega-reset}), $\w{q}$ of the ODE integrator is also
rescaled to unit length.

Equation~(\ref{eq:naiveResult}) can also be derived with the formal
procedure introduced in Section \ref{sec:genDeriv}. This derivation
will highlight an ambiguity not visible in Eq.~(\ref{eq:naiveResult}),
and will also result in the control parameters for scaling and
translation. We start with
\begin{equation}
\w{\bar{x}}=a\,\w{qxq^{*}+T}
\end{equation}
where $\w{\bar{x},x,q,T}$ are quaternions and $a\in\mathbb{R}$. All
vector quantities are now treated as quaternions via the
identification map $\w{v}=(0,\vec{v})$. We now
perturb $a\rightarrow a+ \delta a, \ \w{T} \rightarrow \w{T} +
\w{\delta T}, \ \w{q} \rightarrow \w{q}\left(1+\frac{\w{\delta
      q}}{2}\right)$. The $\w{q}$ - perturbation will result in
vectors $\vec{v}$ being mapped to
\begin{equation}
\w{v}' = \w{q}\left(1+\frac{\w{\delta q}}{2}\right) \w{v}
\left(1-\frac{\w{\delta q}}{2}\right)\w{q}^{*} = \w{qwq^{*}}  
\end{equation}
where $\w{w} \equiv \left(1+\frac{\w{\delta q}}{2}\right) \w{v}
\left(1-\frac{\w{\delta q}}{2}\right)$. This shows that the imaginary
part $\vec{\delta q}$ of $\w{\delta q} =(0,\vec{\delta q})$
represents a rotation in \emph{grid} coordinates.

The quaternion version of  Eq. (\ref{eq:generalSys}) is:
\begin{align}
  \label{eq:mainSystem}
  &a\w{qx_{i}q^{*}+T}= \nonumber \\
  &(a+\delta a) \w{q}\left(1+\frac{\w{\delta
      q}}{2}\right)\w{c_{i}}\left(1-\frac{\w{\delta q}}{2}\right)\w{q}^{*} +
  \w{T+\delta T}.
\end{align}
with $i=A,B$. Because the real part of Eqs.~(\ref{eq:mainSystem}) are
trivially satisfied, Eqs.~(\ref{eq:mainSystem}) represent six
equations, three each for black hole A and for black hole B. We seek
to solve Eqs.~(\ref{eq:mainSystem}) for the unknowns $\delta a,\
\w{\delta T}\!=\! (0,\vec{\delta T}),$ and $\w{\delta q}
\!=\!(0,\vec{\delta q})$. Because $\w{\delta T}$ and $\w{\delta q}$
have three components each, we have in total seven unknowns. The
additional degree of freedom arises because the rotation around
$\vec{C}$ is not yet fixed. Recall that in the Euler angle
representation, we remove this degree of freedom by setting $\phi=0$,
c.f. Eq.~(\ref{eq:PitchYawRotation}). Expanding
Eq.~(\ref{eq:mainSystem}) to linear order in the perturbations and
subtracting the equation for black hole B from that for black hole A,
it is straightforward to show that
\begin{equation}
\label{eq:scaleRes}
\delta a = \left(\frac{\vec{X} \cdot \vec{C}
  }{\norm{\vec{C}}^{2} } -1 \right)a,
\end{equation}
\begin{equation}
\label{eq:rotRes}
\vec{\delta q} = \frac{\vec{C}
\times \vec{X}}{\norm{\vec{C}}^{2}} + \alpha \vec{C},
\end{equation}
and 
\begin{equation}
\label{eq:transRes}
(0,\vec{\delta T})=
a\w{q}\left(\w{x}_{A}-\w{c}_{A} - \w{\delta q} \wedge \w{c}_{A}
  -\frac{\delta a }{a} \w{c}_{A} \right)\w{q}^{*}.
\end{equation}
In Eq.~(\ref{eq:transRes}, $\w{\delta q} \wedge \w{c}_{A} \equiv
(0,\vec{\delta q} \times \vec{c}_{A})$ and $\w{\delta q},\ \delta a$
are to be substituted from Eq.~(\ref{eq:scaleRes}, \ref{eq:rotRes}).
The parameter $\alpha$ in Eq.~(\ref{eq:rotRes}) is undetermined,
reflecting the extra degree of freedom already mentioned after
Eq.~(\ref{eq:mainSystem}). It parameterizes the component of
$\vec\delta q$ parallel to $\vec{C}$; i.e., a rotation about the axis
$\vec{C}$ connecting the two excision spheres. We shall choose it to
minimize the overall rotation $\norm{\vec{\delta q}}$:
\begin{equation}
\label{eq:alphaCond}
\alpha = 0.
\end{equation} 
With this choice Eq.~(\ref{eq:rotRes}) simplifies to
Eq.~(\ref{eq:naiveResult}). The choice $\alpha=0$ is equivalent to the
minimal rotation frame of Boyle et al.~\cite{Boyle:2011gg}; it
minimizes artificial activity of the control system that is not
connected to the physics of the binary black hole.

\begin{figure}
  \includegraphics[width=0.98\columnwidth]{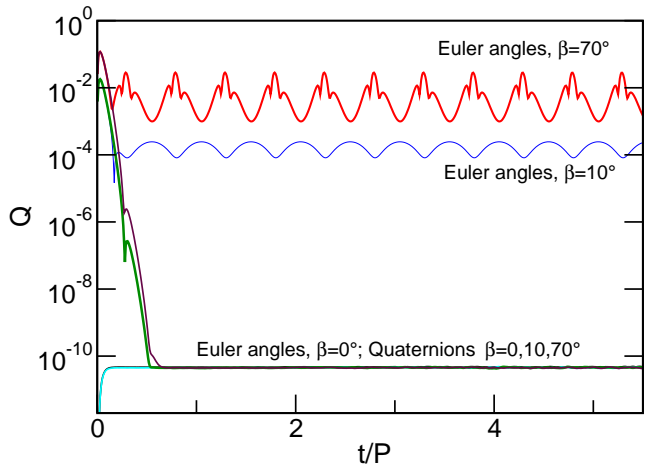}
  \caption{ Newtonian simulations with inclination of the orbital
    plane of angle $\beta=0,10,70$ degrees from the $xy$ plane,
    performed with both control systems. Time is measured in units of
    orbital period.}
  \label{fig:AllNormsN}
\end{figure}

\begin{figure}
  \includegraphics[width=0.98\columnwidth]{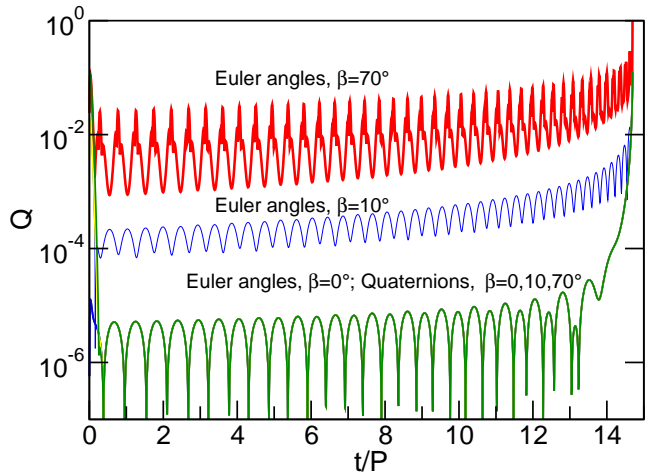}
  \caption{  \label{fig:AllNormsPN}
     Post-Newtonian simulations with inclination of the orbital
    plane angle $\beta=0,10,70$ degrees from the $x-y$
    plane, performed with both control systems. Time is measured in
    units of initial orbital period. The binary is equal mass and
    non-spinning, with the initial coordinate separation of 20. }
\end{figure}


\section{Numerical Results}

\label{sec:Results}

To test our new approach to the rotation control system we begin with
the simplest possible system that still exhibits the desired
behaviour, namely a \emph{Newtonian} circular binary. We consider an
equal mass, non-spinning, circular binary at separation of $20\ M$.
The orbital plane is inclined with respect to the $xy$-plane by angles
$\beta=0,10,70$ degrees. The performance of the control system is
quantified by the magnitude of the control parameters $\norm{Q} \equiv
\sqrt{Q_{i}Q^{i}}$ where the summation extends over all the components
of the control error for rotation, defined by Eqs.~(\ref{eq:QPitch},
\ref{eq:QYaw}) for Euler angles and Eq.~(\ref{eq:rotRes}) for
quaternions. Independent of the inclination $\beta$, we always
initialize the control system as if the binary is in the $xy$-plane.
This is of course only correct for $\beta=0$; for $\beta \ne 0$, the
control system will also have to demonstrate that it can compensate
for an utterly erroneous initialization. Figure \ref{fig:AllNormsN}
shows $\norm{Q}$ for the 3 cases. For $\beta=0^{\circ}$ both control
systems perform very well with an extremely small value of $Q\sim
10^{-11}$. For $\beta \ne 0$, there are initial transients due to the
intentionally wrong initialization of $\lambda^{\mu}$. These
transients decay exponentially on the damping timescale of the control
system; here, $\tau=P/56$ where $P$ is the orbital period. Once the
transients have disappeared for $\beta=10^\circ$, the quaternion
results are unchanged while the Euler-angle control error has
increased by $6$ orders of magnitude. Several periodic sharp features
start to appear. Finally, for $\beta=70^\circ$ the quaternion
$\norm{Q}$ is again at $10^{-11}$ while the Euler-angle $\norm{Q}$
grows by another two orders of magnitude and shows sharp oscillatory
features, which ultimately makes the control system inviable.

Figure~\ref{fig:AllNormsN} foreshadows already the main conclusion of this
work: The Euler-angle approach depends on the plane of the orbit,
and has increasing difficulty in controlling the coordinate mappings
as the orbital plane becomes orthogonal to the xy-plane.
While the Euler-angle control system becomes singular only at
exactly $\beta=90^\circ$, the effects of this singularity are already
clearly visible for $\beta=10^\circ$. In contrast, the quaternion
control system is rotationally invariant, and hence, it controls the
coordinate mapping equally well for any inclination $\beta$.

Next, we turn to a more interesting test that also involves the control
system for the expansion factor $a(t)$. We consider a post-Newtonian
equal mass, non-spinning black hole binary. The relevant
PN equations of motion can be found in \cite{kidder95}.
Figure~\ref{fig:AllNormsPN} displays a set of three runs done with
both control systems, again choosing to tilt the orbit relative to the
$xy$-plane by angles $\beta = 0,10,70$ degrees.

After the initial transients due to intentionally wrong initialization
of the control system, both of the control systems handle the
$\beta=0$ case equally well with $\norm{Q}\sim 10^{-5}$ showing regular
oscillations due to a small eccentricity of the orbit. When
$\beta=10^\circ$, the quaternion control system performs in exactly the
same fashion as for $\beta=0^\circ$. The Euler-angle system, on the
other hand, begins to struggle: the amplitude of $\norm{Q}$ grows by
about two orders of magnitude. The situation grows worse still for Euler
angles when $\beta=70^\circ$, where the control error increases by another
two orders of magnitude and sharp features appear.

Meanwhile, the curves corresponding to the quaternion control system
show exactly (to within numerical accuracy) the same value of
$\norm{Q}$ for all inclinations. This is exactly what we expect from a
rotationally invariant control system - the orientation of the orbital
plane is irrelevant.
\begin{table*}

  \begin{tabular}{|c|c|c|c|c|c|c|}
   \hline
   Name & q & $\vec{\chi}_{1}$ & $\vec{\chi}_{2}$ & $D_0/M$ & $\dot{a}_{0}$ &  $M\Omega_0$    \\
   \hline
   d11.68q2.5 & 2.5  &   (0.000, 0.575, -0.556) & (0.000,
   0.360,-0.347) & 11.68 & -0.000290589649 & 0.02264246       \\
   d12q2.5    & 2.5  &   (0.000, 0.410, -0.287) & (0, 0, 0) & 12    & -0.000108923113 & 0.02180603     \\
   d14.5q1.5  & 1.5  &   (0.000,0.285, 0.093)   & (0, 0, 0)  & 14.5 & -0.000016947638 & 0.01664958     \\
\hline
\end{tabular}
\caption{The initial conditions used for the numerical relativity
  runs. Given are the mass ratio $q=m_1/m_2$, the dimensionless
  spin-vectors $\vec\chi_1$ and $\vec\chi_2 $, the initial separation
  $D_0$,  initial radial velocity
  $\dot{a}_{0}$, initial orbital frequency $\Omega_0$. The initial orbital
  angular momentum is in the $\hat{z}$ direction and the line
  connecting the two black holes is parallel to the $x$-axis.  
  \vspace{0.5em}}
  \label{tab:runs}

\end{table*}

\begin{figure}
  
  \includegraphics[scale=0.33]{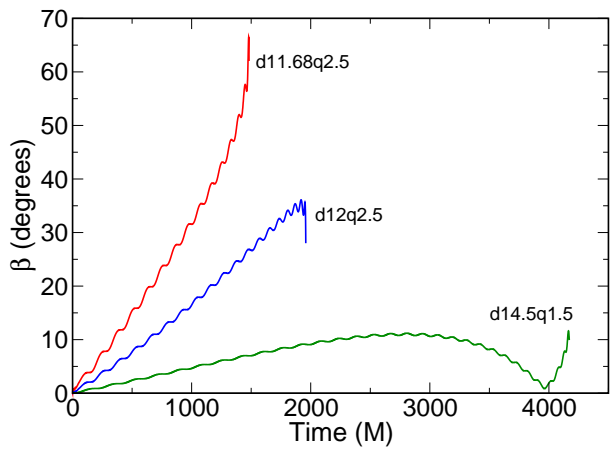}
  \caption{The inclination angle, $\beta$ for the three systems under study. }
  \label{fig:inclination}
\end{figure}
Finally, we test the quaternion-based control system using its main
application: simulations of precessing binary black hole systems in full
numerical relativity. Quite generally, this precession may cause the
orbital plane to rotate by 90 or more degrees with respect to the
initial conditions. The behaviour of the system depends on mass ratio
and the two spin vectors. We choose a set of three simulations to be
evolved using full numerical relativity that exhibits mild to
significant precession. Table \ref{tab:runs} summarizes the initial
conditions. The initial data was constructed from the superposition of
two Kerr-Schild metrics for the conformal metric as in
\cite{Lovelace2008} (so called SKS initial data). The eccentricity has
been removed by an iterative process \cite{Buonanno:2010yk}, so that
the final eccentricities for all three cases are a few $\times
10^{-4}$. 

\begin{figure}
  \begin{tabular}{cc}
    \includegraphics[width=0.5\linewidth]{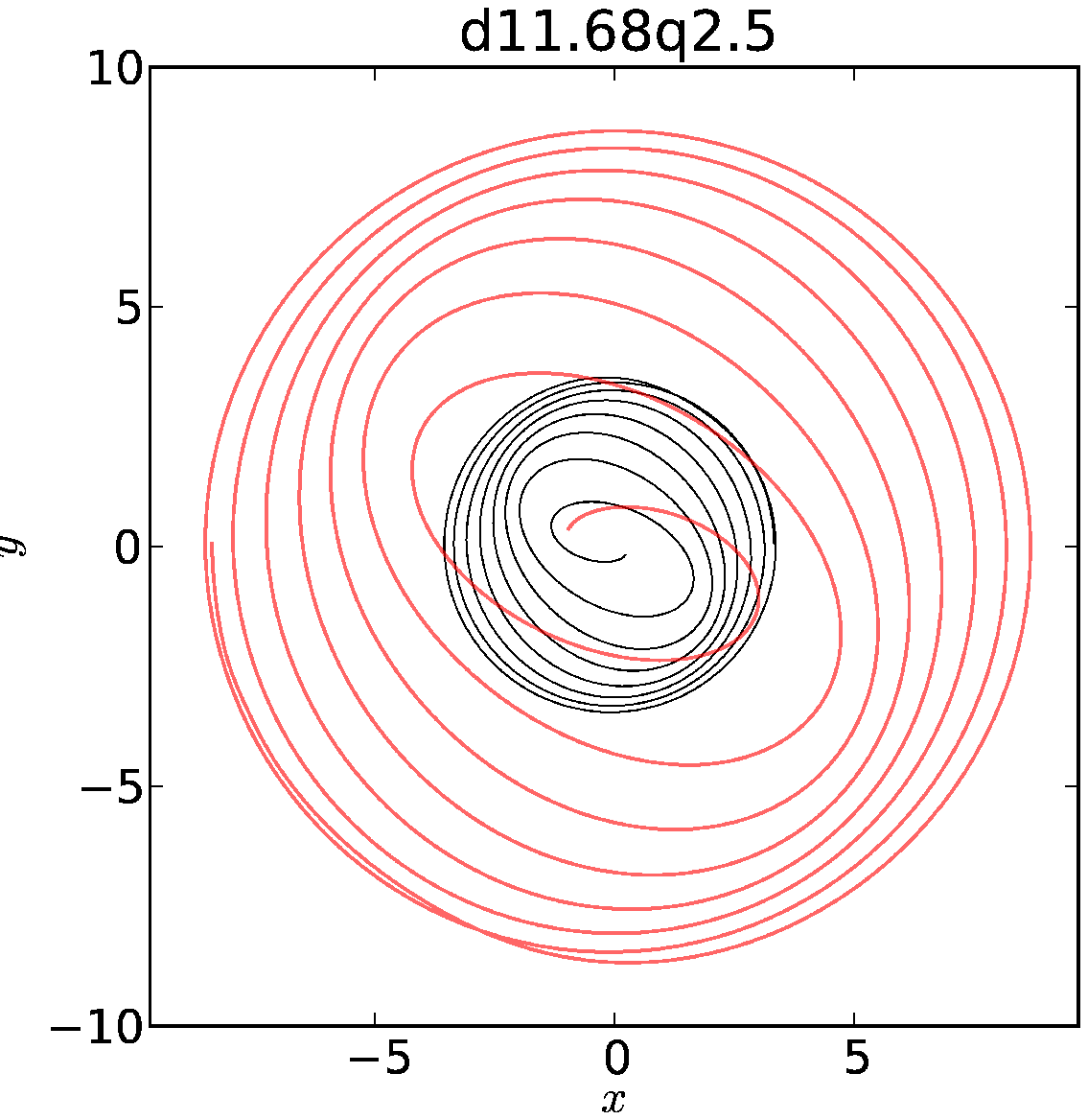}
    &  \includegraphics[width=0.5\linewidth]{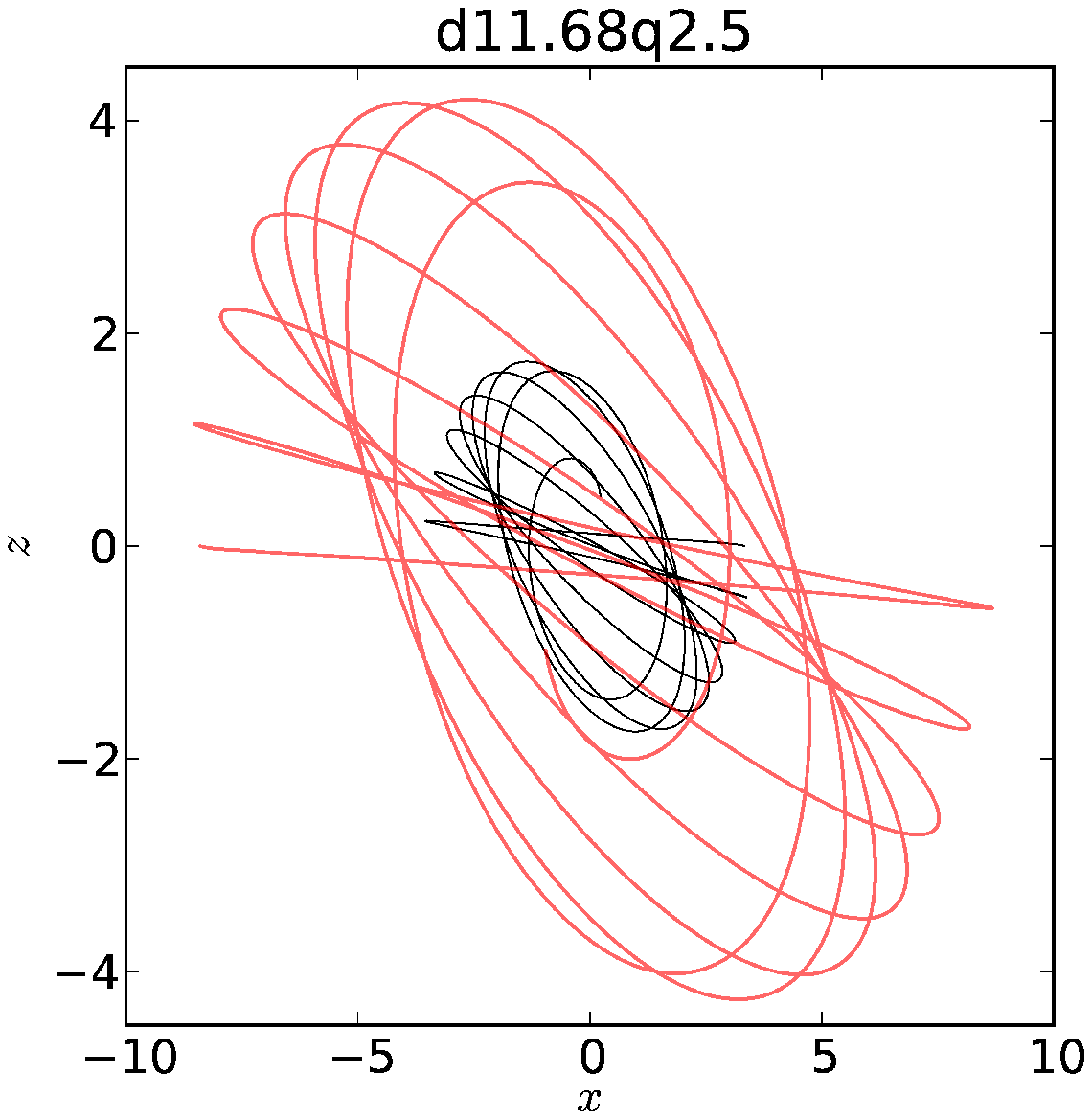} \\
     \includegraphics[width=0.5\linewidth]{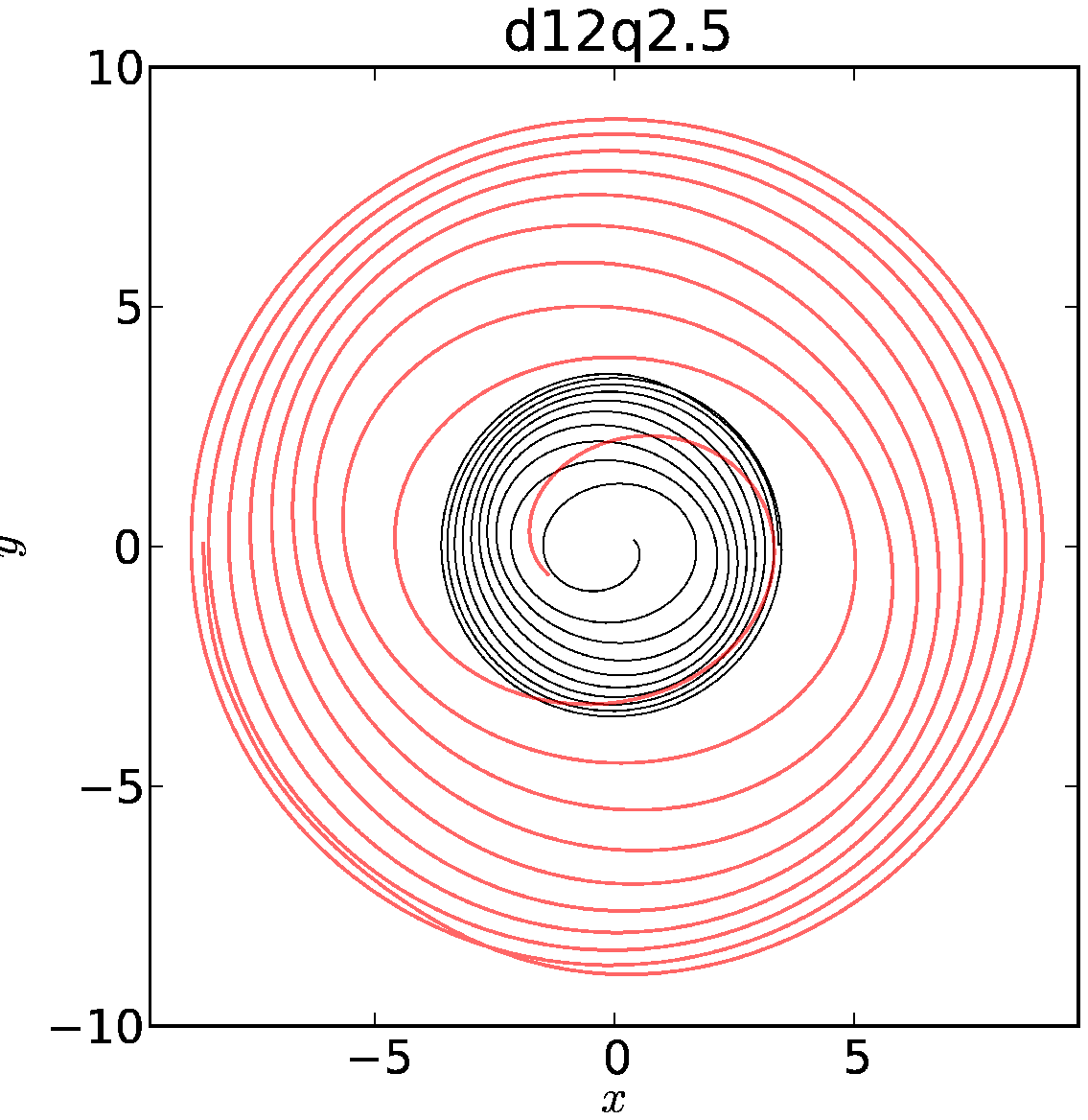}
    &  \includegraphics[width=0.5\linewidth]{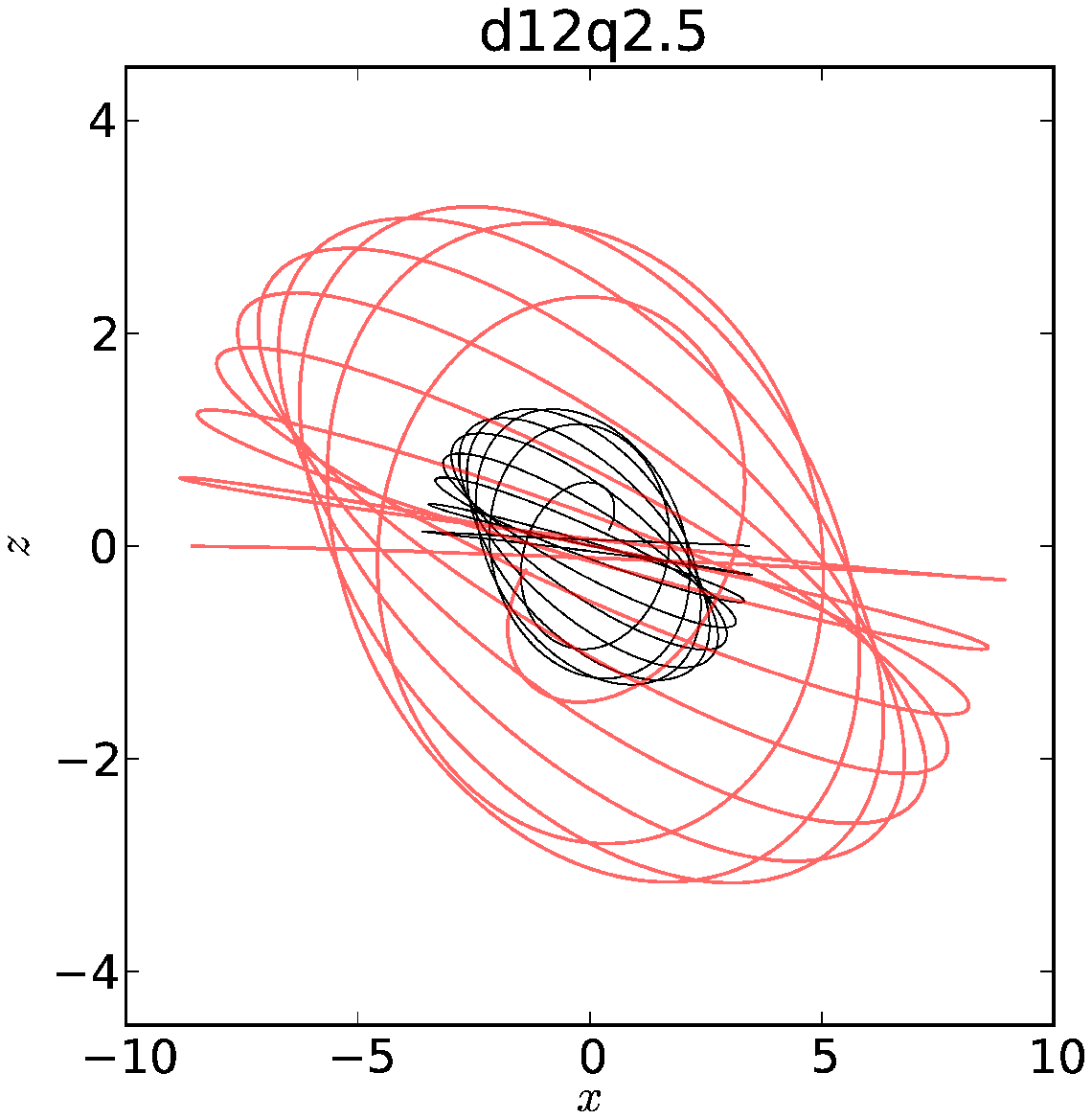} \\
 \includegraphics[width=0.5\linewidth]{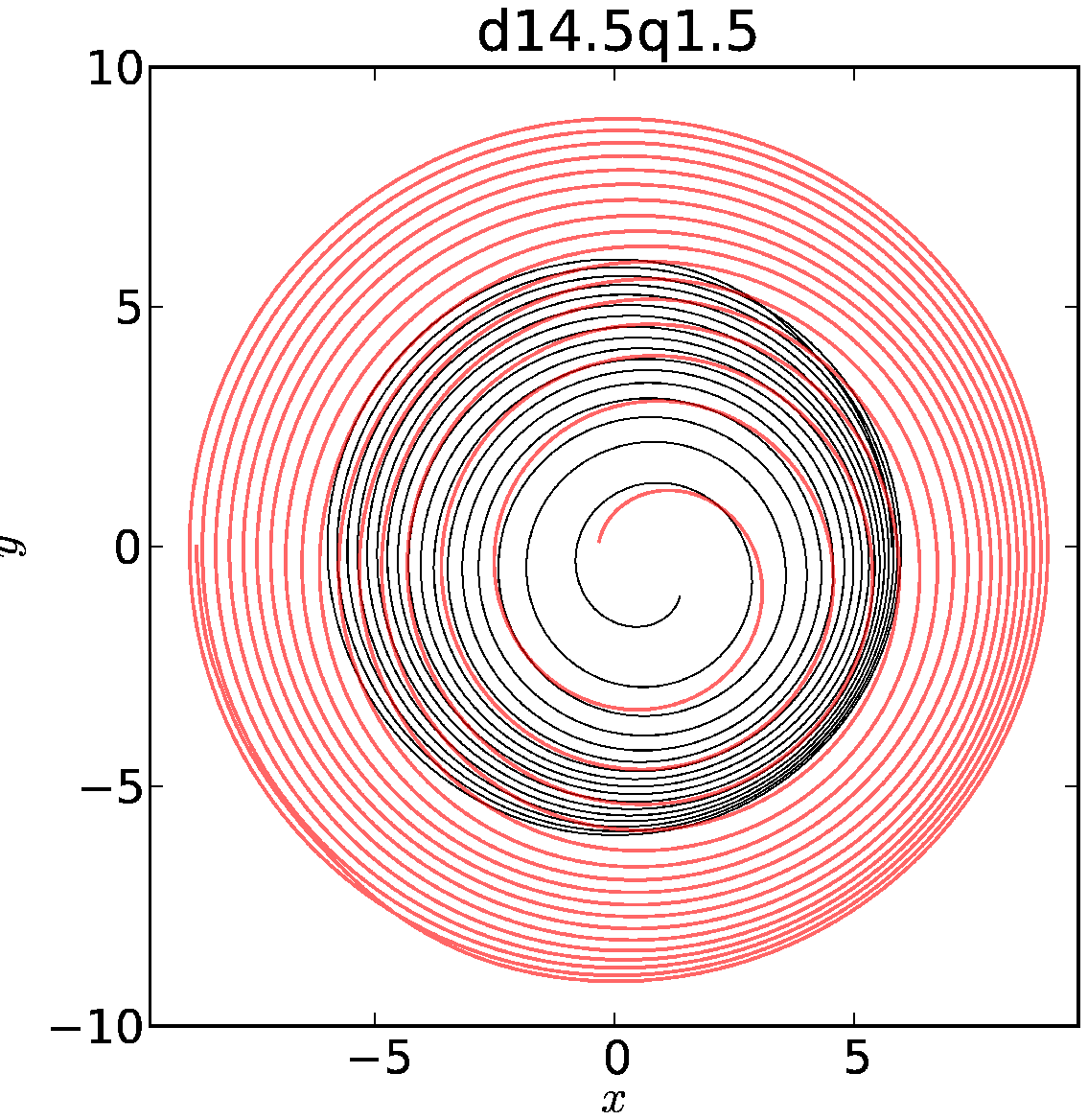}
    &  \includegraphics[width=0.5\linewidth]{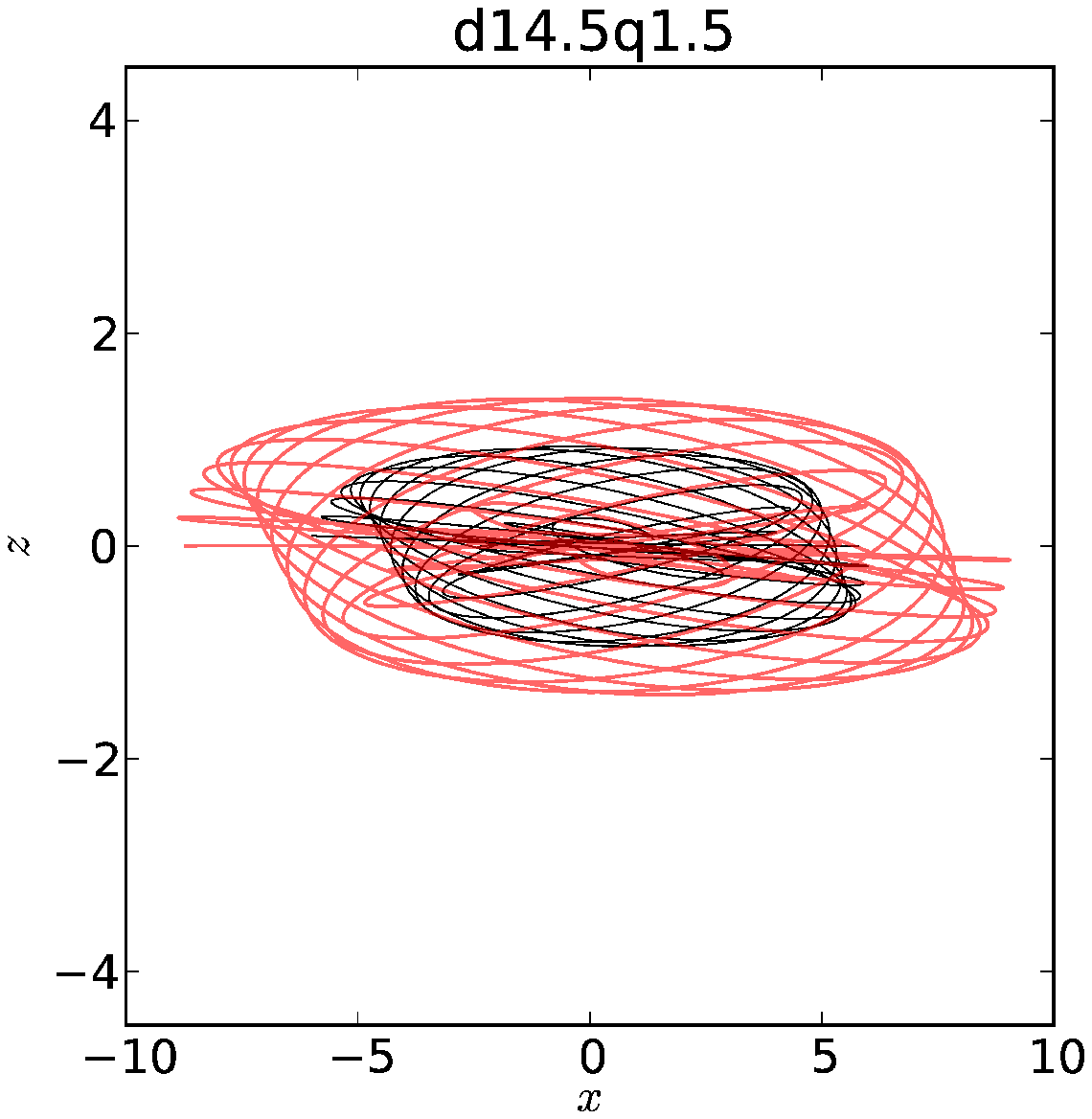}

  \end{tabular}

 \caption{The trajectories of the centers of the apparent horizons of
   the black holes in inertial coordinates for the 3 simulations. Top
   to bottom:   d11.68q2.5, d12q2.5, d14.5q1.5. The left panels show
   the projection onto the xy plane and the right, the xz plane}
  \label{fig:trajs}
\end{figure}

Figure \ref{fig:inclination} shows the inclination angle $\beta \equiv
\arccos(\Omega_{z}/|\vec{\Omega}|)$ which measures the angle between
the normal to the instantaneous orbital plane and the initial
direction of the normal, which is by convention in the $z$-direction.
The high-frequency oscillatory features are due to the nutation of the
orbital angular momentum, while the secular evolution is due to
precession. Notice that the d14.5q1.5 run completes a full precession
cycle at $t=4000M$. The curves end when the black holes merge. The
maximum inclination angles are similar to those used for the
Post-Newtonian evolutions above. Figure \ref{fig:trajs} shows the
trajectories of the black holes in inertial coordinates.

\begin{figure}
 
   \includegraphics[scale=0.33]{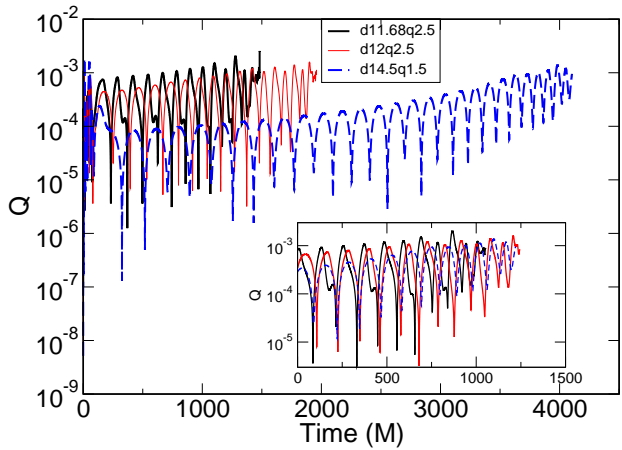}
  \caption{Three full NR simulations performed with the quaternion
    control systems. The initial conditions are listed in Table \ref{tab:runs}}
  \label{fig:AllNorms}
\end{figure}

Figure \ref{fig:AllNorms} presents $\norm{Q}$ for the three runs done
with the quaternion control system. In the main panel of Figure
\ref{fig:AllNorms} it is difficult to compare the $\norm{Q}$ of
different runs because of the difference in orbital frequency. The
inset shows the same $\norm{Q}$ residuals for the three simulations
but timeshifted such that the orbital frequency of $M\Omega = 0.025$
occurs at $t=0$. As one can see the control error norms lie very close
to each other, and exhibit qualitatively similar oscillations with
virtually no sharp features.The remaining differences in behaviour are
associated mostly with the different eccentricities as well as masses
and spins. The growth of the control parameters with time is caused by
the more rapid inspiral towards merger. Our numerical experiments
demonstrate that the quaternion approach is indeed suitable for
simulating arbitrarily precessing configurations.

\section{Discussion}
\label{sec:Discussion}

Simulating precessing binaries poses a challenge for excision-based
numerical techniques. This challenge is resolved here by developing
coordinate mappings which make the black holes be at rest in grid
coordinates. This transformation is dynamically controlled by a
feedback control system since the trajectories of the black holes are
not known in advance. In the most general case this map involves a
rotation, and while Euler angle parametrization works well for mildly
precessing binaries, it exhibits coordinate singularities for polar
orbits which leads to the breakdown of the simulation. To rectify the
situation, we have created a control system that represents rotations
using quaternions. Quaternions do not suffer from coordinate
singularities and work for generically precessing systems. The
quaternion-based control system is able to successfully perform fully
general relativistic simulations of highly-precessing binaries,
allowing the investigation strongly precessing binary black holes and
broadening the range of parameter space that can be explored. The
techniques developed here have already been utilized in the
simulations presented in ~\cite{PhysRevD.87.084006,Mroue:ManyMergersPRL}

The quaternion control system, as described and developed above, is
related to the minimal rotation frame~\cite{Boyle:2011gg} (see
also~\cite{Buonanno:2002fy}). For the control-system developed here,
as in the minimal rotation frame, a preferred {\em axis} exists (the
line connecting the two black holes vs. the instantaneous preferred
emission axis of the gravitational waves). In both cases, the rotation
{\em about} this axis is not a priori determined. And in both cases,
this rotation is chosen such that the instantaneous rotation frequency
of the rotating frame is minimal. In the present context, this
condition is imposed by Eq.~(\ref{eq:alphaCond}).

As a useful byproduct of the quaternion control system, one obtains an
accurate estimate of the orbital frequency and the orbital phase
during the numerical run without the need for any post-processing.
The $\w{\Omega}$ in Eq.~(\ref{eq:qderiv}) is the instantaneous rotation
frequency of the grid frame relative to the inertial frame, given in
components of the grid frame. Converting to the inertial frame, 

\begin{equation}
  \label{eq:omegaTrans}
  (0,\vec{\omega})=\w{q\Omega q^{*}}.
\end{equation}

\begin{figure}

   \includegraphics[scale=0.33]{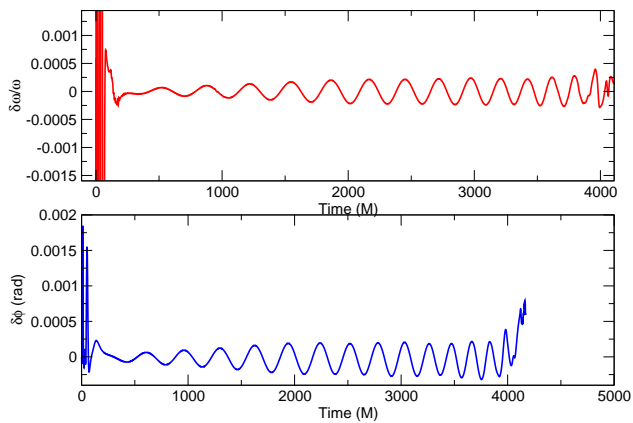}
   \caption{Top: the fractional difference in orbital frequency
     estimated from quaternions and from trajectories. Bottom: the
     orbital phase difference in radians. The data is from the run
     d14.5q1.5 }
  \label{fig:OmegaComp}
\end{figure}

If the control system were perfect - i.e. if $Q \equiv 0$ - then
$\vec{\omega}$ given by Eq.~(\ref{eq:omegaTrans}) would be the
instantaneous orbital frequency. Because $Q\ne 0$,
Eq.~(\ref{eq:omegaTrans}) only gives an approximate orbital frequency,
albeit a very good one: The upper panel of Fig.~\ref{fig:OmegaComp}
shows the fractional difference between $|\vec{\omega}|$ from
Eq.~(\ref{eq:omegaTrans}) and the exact numerical orbital frequency
obtained by post-processing. The difference oscillates around zero
with relative amplitude of $1\times10^{-3}$. It is also
straightforward to integrate

\begin{equation}
  \label{eq:phaseEv}
   \dot{\phi} =|\vec{\Omega}|=|\vec{\omega}|
\end{equation}
to obtain the orbital phase of the precessing binary. In practice, we
add Eq.~(\ref{eq:phaseEv}) to the set of ordinary differential
equations Eq.~(\ref{eq:qderiv}) that are integrated to obtain the
rotation quaternion $\w{q}(t)$. The difference between the orbital
phase from the control system Eq.~(\ref{eq:phaseEv}) and the exact
orbital phase from the BH trajectories is shown in the lower panel of
Fig.~\ref{fig:OmegaComp}. The difference is $\sim 10^{-4}$ radians
until merger, during an inspiral lasting 105 radians. Incidentally,
this again demonstrates that our control system works exactly as
expected.

\acknowledgments

We thank Mark Scheel for assistance in implementing the quaternion
control system in SpEC, and Mark Scheel and Bela Szilagyi for
assistance with running the mergers of the black hole binaries
discussed in Sec.~\ref{sec:Results}.
We gratefully acknowledge support from NSERC of Canada, from the Canada
Research Chairs Program, and from the Canadian Institute for Advanced
Research.  We further gratefully acknowledge support from the Sherman
Fairchild Foundation, 
NSF Grants No. PHY-0969111 and No. PHY-1005426 
and NASA Grant No. NNX09AF96G at Cornell.
Calculations were performed at the GPC supercomputer at the SciNet HPC
Consortium; SciNet is funded by: the Canada Foundation for Innovation (CFI)
under the auspices of Compute Canada; the Government of Ontario;
Ontario Research Fund (ORF) -- Research Excellence; and the University of
Toronto. Further computations were performed on the Caltech compute
cluster Zwicky, which was funded by the Sherman Fairchild Foundation
and the NSF MRI-R2 grant No. PHY-0960291, and on the Gravity compute
cluster funded by CFI, ORF, and the University of Toronto.

\section*{Appendix}
\subsection{Rotation control parameters in matrix notation}

 The underlying idea for the rotational control system that we
  have described in section \ref{sec:constr} is independent of the use of
  quaternions to represent rotations. For example, one could have used
  infinitesimal rotation matrices to achieve the same goal. Below is
  the demonstration of the same derivation as in section
  \ref{sec:constr} but now in terms of a rotation matrix $R$. We
  start with the following version of Eq.~(\ref{eq:generalSys}):
\begin{equation}
\label{eq:matrixSys}
 aR\vec{x}_{i} + \vec{T} = (a+\delta a) R(I+\delta R) \vec{c}_{i} +\vec{T} + \vec{\delta T} 
\end{equation}
where as usual $i=A,\ B$, $ \vec{x}_{i}, \vec{c}_{i}, \vec{T}, \
\vec{\delta T} \in \mathbb{R}^{3} $, $ I $ is the identity matrix, $
a,\ \delta a \in \mathbb{R} $ and $ R,\ \delta R \in M_{3\times 3} $ .
Once more we seek to solve this system of six equations for the
unknowns $\delta a$, $\vec{\delta T}$, and $\delta R$. Note that since
$\delta R$ is an infinitesimal rotation matrix, it is skew symmetric
and thus has three independent components, for a total of seven
unknowns. Expanding Eq.~(\ref{eq:matrixSys}) to first order in
perturbation and subtracting the equation for black hole B from that
of black hole A one can show that:
\begin{equation}
\label{eq:scaleResM}
\delta a = \left(\frac{\vec{X} \cdot \vec{C}
  }{\norm{\vec{C}}^{2} } -1 \right)a,
\end{equation}
\begin{equation}
\label{eq:rotResM}
\delta R_{ij}=\epsilon_{ijk}\delta\phi^{j},\ \vec{\delta \phi} = \frac{\vec{C}
  \times \vec{X}}{\norm{\vec{C}}^{2}} + \alpha \vec{C},
\end{equation}
\begin{equation}
\label{eq:transResM}
\vec{\delta T}= aR\left(\vec{x}_{A}-\vec{c}_{A}-\vec{\delta \phi} \times
\vec{c}_{A} - \frac{\delta a}{a}\vec{c}_{A}\right).
\end{equation}
These results match exactly Eqs.~(\ref{eq:scaleRes})-(\ref{eq:transRes}). Thus we see that indeed
  infinitesimal rotation matrices could have been used to represent
  rotation. We selected quaternions for our work primarily for
  numerical reasons, the main being the ease of correcting numerical
  drift from a rotation, which for quaternions amounts to a simple
  renormalization. 

\hbadness 10000\relax
\bibliography{References}

\begin{thebibliography}{53}%
\makeatletter
\providecommand \@ifxundefined [1]{%
 \@ifx{#1\undefined}
}%
\providecommand \@ifnum [1]{%
 \ifnum #1\expandafter \@firstoftwo
 \else \expandafter \@secondoftwo
 \fi
}%
\providecommand \@ifx [1]{%
 \ifx #1\expandafter \@firstoftwo
 \else \expandafter \@secondoftwo
 \fi
}%
\providecommand \natexlab [1]{#1}%
\providecommand \enquote  [1]{``#1''}%
\providecommand \bibnamefont  [1]{#1}%
\providecommand \bibfnamefont [1]{#1}%
\providecommand \citenamefont [1]{#1}%
\providecommand \href@noop [0]{\@secondoftwo}%
\providecommand \href [0]{\begingroup \@sanitize@url \@href}%
\providecommand \@href[1]{\@@startlink{#1}\@@href}%
\providecommand \@@href[1]{\endgroup#1\@@endlink}%
\providecommand \@sanitize@url [0]{\catcode `\\12\catcode `\$12\catcode
  `\&12\catcode `\#12\catcode `\^12\catcode `\_12\catcode `\%12\relax}%
\providecommand \@@startlink[1]{}%
\providecommand \@@endlink[0]{}%
\providecommand \url  [0]{\begingroup\@sanitize@url \@url }%
\providecommand \@url [1]{\endgroup\@href {#1}{\urlprefix }}%
\providecommand \urlprefix  [0]{URL }%
\providecommand \Eprint [0]{\href }%
\providecommand \doibase [0]{http://dx.doi.org/}%
\providecommand \selectlanguage [0]{\@gobble}%
\providecommand \bibinfo  [0]{\@secondoftwo}%
\providecommand \bibfield  [0]{\@secondoftwo}%
\providecommand \translation [1]{[#1]}%
\providecommand \BibitemOpen [0]{}%
\providecommand \bibitemStop [0]{}%
\providecommand \bibitemNoStop [0]{.\EOS\space}%
\providecommand \EOS [0]{\spacefactor3000\relax}%
\providecommand \BibitemShut  [1]{\csname bibitem#1\endcsname}%
\let\auto@bib@innerbib\@empty
\bibitem [{\citenamefont {Aasi}\ \emph {et~al.}(2013)\citenamefont {Aasi} \emph
  {et~al.}}]{Aasi:2013wya}%
  \BibitemOpen
  \bibfield  {author} {\bibinfo {author} {\bibfnamefont {J.}~\bibnamefont
  {Aasi}} \emph {et~al.} (\bibinfo {collaboration} {LIGO Scientific
  Collaboration, Virgo Collaboration}),\ }\href@noop {} {\  (\bibinfo {year}
  {2013})},\ \Eprint {http://arxiv.org/abs/1304.0670} {arXiv:1304.0670 [gr-qc]}
  \BibitemShut {NoStop}%
\bibitem [{\citenamefont {Abadie}\ \emph {et~al.}(2010)\citenamefont {Abadie}
  \emph {et~al.}}]{Abadie:2010cfa}%
  \BibitemOpen
  \bibfield  {author} {\bibinfo {author} {\bibfnamefont {J.}~\bibnamefont
  {Abadie}} \emph {et~al.} (\bibinfo {collaboration} {LIGO Scientific}),\
  }\href {\doibase 10.1088/0264-9381/27/17/173001} {\bibfield  {journal}
  {\bibinfo  {journal} {Class. Quant. Grav.}\ }\textbf {\bibinfo {volume}
  {27}},\ \bibinfo {pages} {173001} (\bibinfo {year} {2010})},\ \Eprint
  {http://arxiv.org/abs/1003.2480} {arXiv:1003.2480 [Unknown]} \BibitemShut
  {NoStop}%
\bibitem [{\citenamefont {Finn}(1992)}]{Finn1992}%
  \BibitemOpen
  \bibfield  {author} {\bibinfo {author} {\bibfnamefont {L.~S.}\ \bibnamefont
  {Finn}},\ }\href@noop {} {\bibfield  {journal} {\bibinfo  {journal} {Phys.\
  Rev.\ D}\ }\textbf {\bibinfo {volume} {46}},\ \bibinfo {pages} {5236}
  (\bibinfo {year} {1992})}\BibitemShut {NoStop}%
\bibitem [{\citenamefont {Finn}\ and\ \citenamefont
  {Chernoff}(1993)}]{Finn1993}%
  \BibitemOpen
  \bibfield  {author} {\bibinfo {author} {\bibfnamefont {L.~S.}\ \bibnamefont
  {Finn}}\ and\ \bibinfo {author} {\bibfnamefont {D.~F.}\ \bibnamefont
  {Chernoff}},\ }\href@noop {} {\bibfield  {journal} {\bibinfo  {journal}
  {Phys.\ Rev.\ D}\ }\textbf {\bibinfo {volume} {47}},\ \bibinfo {pages} {2198}
  (\bibinfo {year} {1993})}\BibitemShut {NoStop}%
\bibitem [{\citenamefont {Ohme}(2012)}]{Ohme:2011rm}%
  \BibitemOpen
  \bibfield  {author} {\bibinfo {author} {\bibfnamefont {F.}~\bibnamefont
  {Ohme}},\ }\href@noop {} {\bibfield  {journal} {\bibinfo  {journal}
  {Class.Quant.Grav.}\ }\textbf {\bibinfo {volume} {29}},\ \bibinfo {pages}
  {124002} (\bibinfo {year} {2012})},\ \Eprint {http://arxiv.org/abs/1111.3737}
  {arXiv:1111.3737 [gr-qc]} \BibitemShut {NoStop}%
\bibitem [{\citenamefont {Pretorius}(2005)}]{Pretorius2005a}%
  \BibitemOpen
  \bibfield  {author} {\bibinfo {author} {\bibfnamefont {F.}~\bibnamefont
  {Pretorius}},\ }\href {\doibase 10.1103/PhysRevLett.95.121101} {\bibfield
  {journal} {\bibinfo  {journal} {Phys.Rev.Lett.}\ }\textbf {\bibinfo {volume}
  {95}},\ \bibinfo {pages} {121101} (\bibinfo {year} {2005})},\ \Eprint
  {http://arxiv.org/abs/gr-qc/0507014} {arXiv:gr-qc/0507014 [gr-qc]}
  \BibitemShut {NoStop}%
\bibitem [{\citenamefont {Hinder}(2010)}]{Hinder:2010vn}%
  \BibitemOpen
  \bibfield  {author} {\bibinfo {author} {\bibfnamefont {I.}~\bibnamefont
  {Hinder}},\ }\href {\doibase 10.1088/0264-9381/27/11/114004} {\bibfield
  {journal} {\bibinfo  {journal} {Class. Quant. Grav.}\ }\textbf {\bibinfo
  {volume} {27}},\ \bibinfo {pages} {114004} (\bibinfo {year} {2010})},\
  \Eprint {http://arxiv.org/abs/1001.5161} {arXiv:1001.5161 [gr-qc]}
  \BibitemShut {NoStop}%
\bibitem [{\citenamefont {Pfeiffer}(2012)}]{Pfeiffer:2012pc}%
  \BibitemOpen
  \bibfield  {author} {\bibinfo {author} {\bibfnamefont {H.~P.}\ \bibnamefont
  {Pfeiffer}},\ }\href {\doibase 10.1088/0264-9381/29/12/124004} {\bibfield
  {journal} {\bibinfo  {journal} {Class.Quant.Grav.}\ }\textbf {\bibinfo
  {volume} {29}},\ \bibinfo {pages} {124004} (\bibinfo {year} {2012})},\
  \Eprint {http://arxiv.org/abs/1203.5166} {arXiv:1203.5166 [gr-qc]}
  \BibitemShut {NoStop}%
\bibitem [{\citenamefont {Blanchet}(2002)}]{lrr-2002-3}%
  \BibitemOpen
  \bibfield  {author} {\bibinfo {author} {\bibfnamefont {L.}~\bibnamefont
  {Blanchet}},\ }\href {http://www.livingreviews.org/lrr-2002-3} {\bibfield
  {journal} {\bibinfo  {journal} {Living Reviews in Relativity}\ }\textbf
  {\bibinfo {volume} {5}} (\bibinfo {year} {2002})}\BibitemShut {NoStop}%
\bibitem [{\citenamefont {Blanchet}(2011)}]{Blanchet:2009rw}%
  \BibitemOpen
  \bibfield  {author} {\bibinfo {author} {\bibfnamefont {L.}~\bibnamefont
  {Blanchet}},\ }\href@noop {} {\bibfield  {journal} {\bibinfo  {journal}
  {Fundam.Theor.Phys.}\ }\textbf {\bibinfo {volume} {162}},\ \bibinfo {pages}
  {125} (\bibinfo {year} {2011})},\ \Eprint {http://arxiv.org/abs/0907.3596}
  {arXiv:0907.3596 [gr-qc]} \BibitemShut {NoStop}%
\bibitem [{\citenamefont {Buonanno}\ and\ \citenamefont
  {Damour}(1999)}]{Buonanno99}%
  \BibitemOpen
  \bibfield  {author} {\bibinfo {author} {\bibfnamefont {A.}~\bibnamefont
  {Buonanno}}\ and\ \bibinfo {author} {\bibfnamefont {T.}~\bibnamefont
  {Damour}},\ }\href {\doibase 10.1103/PhysRevD.59.084006} {\bibfield
  {journal} {\bibinfo  {journal} {Phys.Rev.}\ }\textbf {\bibinfo {volume}
  {D59}},\ \bibinfo {pages} {084006} (\bibinfo {year} {1999})},\ \Eprint
  {http://arxiv.org/abs/gr-qc/9811091} {arXiv:gr-qc/9811091 [gr-qc]}
  \BibitemShut {NoStop}%
\bibitem [{\citenamefont {Damour}(2001)}]{Damour01c}%
  \BibitemOpen
  \bibfield  {author} {\bibinfo {author} {\bibfnamefont {T.}~\bibnamefont
  {Damour}},\ }\href {\doibase 10.1103/PhysRevD.64.124013} {\bibfield
  {journal} {\bibinfo  {journal} {Phys.Rev.}\ }\textbf {\bibinfo {volume}
  {D64}},\ \bibinfo {pages} {124013} (\bibinfo {year} {2001})},\ \Eprint
  {http://arxiv.org/abs/gr-qc/0103018} {arXiv:gr-qc/0103018 [gr-qc]}
  \BibitemShut {NoStop}%
\bibitem [{\citenamefont {Damour}\ \emph {et~al.}(2009)\citenamefont {Damour},
  \citenamefont {Iyer},\ and\ \citenamefont {Nagar}}]{DIN}%
  \BibitemOpen
  \bibfield  {author} {\bibinfo {author} {\bibfnamefont {T.}~\bibnamefont
  {Damour}}, \bibinfo {author} {\bibfnamefont {B.~R.}\ \bibnamefont {Iyer}}, \
  and\ \bibinfo {author} {\bibfnamefont {A.}~\bibnamefont {Nagar}},\ }\href
  {\doibase 10.1103/PhysRevD.79.064004} {\bibfield  {journal} {\bibinfo
  {journal} {Phys.Rev.}\ }\textbf {\bibinfo {volume} {D79}},\ \bibinfo {pages}
  {064004} (\bibinfo {year} {2009})},\ \Eprint {http://arxiv.org/abs/0811.2069}
  {arXiv:0811.2069 [gr-qc]} \BibitemShut {NoStop}%
\bibitem [{\citenamefont {Damour}(2012)}]{Damour:2012mv}%
  \BibitemOpen
  \bibfield  {author} {\bibinfo {author} {\bibfnamefont {T.}~\bibnamefont
  {Damour}},\ }\href@noop {} {\  (\bibinfo {year} {2012})},\ \Eprint
  {http://arxiv.org/abs/1212.3169} {arXiv:1212.3169 [gr-qc]} \BibitemShut
  {NoStop}%
\bibitem [{\citenamefont {Damour}\ and\ \citenamefont
  {Nagar}(2009)}]{Damour2009a}%
  \BibitemOpen
  \bibfield  {author} {\bibinfo {author} {\bibfnamefont {T.}~\bibnamefont
  {Damour}}\ and\ \bibinfo {author} {\bibfnamefont {A.}~\bibnamefont {Nagar}},\
  }\href {\doibase 10.1103/PhysRevD.79.081503} {\bibfield  {journal} {\bibinfo
  {journal} {Phys.\ Rev.\ D}\ }\textbf {\bibinfo {volume} {79}},\ \bibinfo
  {pages} {081503} (\bibinfo {year} {2009})},\ \Eprint
  {http://arxiv.org/abs/0902.0136} {arXiv:0902.0136 [gr-qc]} \BibitemShut
  {NoStop}%
\bibitem [{\citenamefont {Buonanno}\ \emph {et~al.}(2009)\citenamefont
  {Buonanno}, \citenamefont {Pan}, \citenamefont {Pfeiffer}, \citenamefont
  {Scheel}, \citenamefont {Buchman} \emph {et~al.}}]{Buonanno:2009qa}%
  \BibitemOpen
  \bibfield  {author} {\bibinfo {author} {\bibfnamefont {A.}~\bibnamefont
  {Buonanno}}, \bibinfo {author} {\bibfnamefont {Y.}~\bibnamefont {Pan}},
  \bibinfo {author} {\bibfnamefont {H.~P.}\ \bibnamefont {Pfeiffer}}, \bibinfo
  {author} {\bibfnamefont {M.~A.}\ \bibnamefont {Scheel}}, \bibinfo {author}
  {\bibfnamefont {L.~T.}\ \bibnamefont {Buchman}},  \emph {et~al.},\ }\href
  {\doibase 10.1103/PhysRevD.79.124028} {\bibfield  {journal} {\bibinfo
  {journal} {Phys.\ Rev.\ D}\ }\textbf {\bibinfo {volume} {79}},\ \bibinfo
  {pages} {124028} (\bibinfo {year} {2009})},\ \Eprint
  {http://arxiv.org/abs/0902.0790} {arXiv:0902.0790 [gr-qc]} \BibitemShut
  {NoStop}%
\bibitem [{\citenamefont {Taracchini}\ \emph {et~al.}(2012)\citenamefont
  {Taracchini}, \citenamefont {Buonanno}, \citenamefont {Barausse},
  \citenamefont {Boyle}, \citenamefont {Chu}, \citenamefont {Lovelace},
  \citenamefont {Pfeiffer},\ and\ \citenamefont {Scheel}}]{Taracchini:2012}%
  \BibitemOpen
  \bibfield  {author} {\bibinfo {author} {\bibfnamefont {A.}~\bibnamefont
  {Taracchini}}, \bibinfo {author} {\bibfnamefont {A.}~\bibnamefont
  {Buonanno}}, \bibinfo {author} {\bibfnamefont {E.}~\bibnamefont {Barausse}},
  \bibinfo {author} {\bibfnamefont {M.}~\bibnamefont {Boyle}}, \bibinfo
  {author} {\bibfnamefont {T.}~\bibnamefont {Chu}}, \bibinfo {author}
  {\bibfnamefont {G.}~\bibnamefont {Lovelace}}, \bibinfo {author}
  {\bibfnamefont {H.~P.}\ \bibnamefont {Pfeiffer}}, \ and\ \bibinfo {author}
  {\bibfnamefont {M.~A.}\ \bibnamefont {Scheel}},\ }\href {\doibase
  10.1103/PhysRevD.83.104034} {\bibfield  {journal} {\bibinfo  {journal}
  {Phys.Rev.}\ }\textbf {\bibinfo {volume} {D83}},\ \bibinfo {pages} {104034}
  (\bibinfo {year} {2012})},\ \Eprint {http://arxiv.org/abs/1202.0790}
  {arXiv:1202.0790 [gr-qc]} \BibitemShut {NoStop}%
\bibitem [{\citenamefont {Apostolatos}\ \emph {et~al.}(1994)\citenamefont
  {Apostolatos}, \citenamefont {Cutler}, \citenamefont {Sussman},\ and\
  \citenamefont {Thorne}}]{Apostolatos1994}%
  \BibitemOpen
  \bibfield  {author} {\bibinfo {author} {\bibfnamefont {T.~A.}\ \bibnamefont
  {Apostolatos}}, \bibinfo {author} {\bibfnamefont {C.}~\bibnamefont {Cutler}},
  \bibinfo {author} {\bibfnamefont {G.~J.}\ \bibnamefont {Sussman}}, \ and\
  \bibinfo {author} {\bibfnamefont {K.~S.}\ \bibnamefont {Thorne}},\
  }\href@noop {} {\bibfield  {journal} {\bibinfo  {journal} {Phys.\ Rev.\ D}\
  }\textbf {\bibinfo {volume} {49}},\ \bibinfo {pages} {6274 } (\bibinfo {year}
  {1994})}\BibitemShut {NoStop}%
\bibitem [{\citenamefont {Campanelli}\ \emph {et~al.}(2007)\citenamefont
  {Campanelli}, \citenamefont {Lousto}, \citenamefont {Zlochower},
  \citenamefont {Krishnan},\ and\ \citenamefont {Merritt}}]{Campanelli2007b}%
  \BibitemOpen
  \bibfield  {author} {\bibinfo {author} {\bibfnamefont {M.}~\bibnamefont
  {Campanelli}}, \bibinfo {author} {\bibfnamefont {C.~O.}\ \bibnamefont
  {Lousto}}, \bibinfo {author} {\bibfnamefont {Y.}~\bibnamefont {Zlochower}},
  \bibinfo {author} {\bibfnamefont {B.}~\bibnamefont {Krishnan}}, \ and\
  \bibinfo {author} {\bibfnamefont {D.}~\bibnamefont {Merritt}},\ }\href@noop
  {} {\bibfield  {journal} {\bibinfo  {journal} {Phys.\ Rev.\ D}\ }\textbf
  {\bibinfo {volume} {75}},\ \bibinfo {pages} {064030} (\bibinfo {year}
  {2007})},\ \Eprint {http://arxiv.org/abs/gr-qc/0612076} {gr-qc/0612076}
  \BibitemShut {NoStop}%
\bibitem [{\citenamefont {Campanelli}\ \emph {et~al.}(2009)\citenamefont
  {Campanelli}, \citenamefont {Lousto}, \citenamefont {Nakano},\ and\
  \citenamefont {Zlochower}}]{CampanelliEtal2009}%
  \BibitemOpen
  \bibfield  {author} {\bibinfo {author} {\bibfnamefont {M.}~\bibnamefont
  {Campanelli}}, \bibinfo {author} {\bibfnamefont {C.~O.}\ \bibnamefont
  {Lousto}}, \bibinfo {author} {\bibfnamefont {H.}~\bibnamefont {Nakano}}, \
  and\ \bibinfo {author} {\bibfnamefont {Y.}~\bibnamefont {Zlochower}},\
  }\href@noop {} {\bibfield  {journal} {\bibinfo  {journal} {Phys.\ Rev.\ D}\
  }\textbf {\bibinfo {volume} {79}},\ \bibinfo {pages} {084010} (\bibinfo
  {year} {2009})},\ \Eprint {http://arxiv.org/abs/arXiv:gr-qc/0808.0713}
  {arXiv:gr-qc/0808.0713} \BibitemShut {NoStop}%
\bibitem [{\citenamefont {Sturani}\ \emph
  {et~al.}(2010{\natexlab{a}})\citenamefont {Sturani}, \citenamefont
  {Fischetti}, \citenamefont {Cadonati}, \citenamefont {Guidi}, \citenamefont
  {Healy} \emph {et~al.}}]{Sturani:2010ju}%
  \BibitemOpen
  \bibfield  {author} {\bibinfo {author} {\bibfnamefont {R.}~\bibnamefont
  {Sturani}}, \bibinfo {author} {\bibfnamefont {S.}~\bibnamefont {Fischetti}},
  \bibinfo {author} {\bibfnamefont {L.}~\bibnamefont {Cadonati}}, \bibinfo
  {author} {\bibfnamefont {G.}~\bibnamefont {Guidi}}, \bibinfo {author}
  {\bibfnamefont {J.}~\bibnamefont {Healy}},  \emph {et~al.},\ }\href@noop {}
  {\  (\bibinfo {year} {2010}{\natexlab{a}})},\ \Eprint
  {http://arxiv.org/abs/1012.5172} {arXiv:1012.5172 [gr-qc]} \BibitemShut
  {NoStop}%
\bibitem [{\citenamefont {Sturani}\ \emph
  {et~al.}(2010{\natexlab{b}})\citenamefont {Sturani}, \citenamefont
  {Fischetti}, \citenamefont {Cadonati}, \citenamefont {Guidi}, \citenamefont
  {Healy} \emph {et~al.}}]{Sturani:2010yv}%
  \BibitemOpen
  \bibfield  {author} {\bibinfo {author} {\bibfnamefont {R.}~\bibnamefont
  {Sturani}}, \bibinfo {author} {\bibfnamefont {S.}~\bibnamefont {Fischetti}},
  \bibinfo {author} {\bibfnamefont {L.}~\bibnamefont {Cadonati}}, \bibinfo
  {author} {\bibfnamefont {G.}~\bibnamefont {Guidi}}, \bibinfo {author}
  {\bibfnamefont {J.}~\bibnamefont {Healy}},  \emph {et~al.},\ }\href {\doibase
  10.1088/1742-6596/243/1/012007} {\bibfield  {journal} {\bibinfo  {journal}
  {J.Phys.Conf.Ser.}\ }\textbf {\bibinfo {volume} {243}},\ \bibinfo {pages}
  {012007} (\bibinfo {year} {2010}{\natexlab{b}})},\ \Eprint
  {http://arxiv.org/abs/1005.0551} {arXiv:1005.0551 [gr-qc]} \BibitemShut
  {NoStop}%
\bibitem [{\citenamefont {Schmidt}\ \emph {et~al.}(2011)\citenamefont
  {Schmidt}, \citenamefont {Hannam}, \citenamefont {Husa},\ and\ \citenamefont
  {Ajith}}]{Schmidt2010}%
  \BibitemOpen
  \bibfield  {author} {\bibinfo {author} {\bibfnamefont {P.}~\bibnamefont
  {Schmidt}}, \bibinfo {author} {\bibfnamefont {M.}~\bibnamefont {Hannam}},
  \bibinfo {author} {\bibfnamefont {S.}~\bibnamefont {Husa}}, \ and\ \bibinfo
  {author} {\bibfnamefont {P.}~\bibnamefont {Ajith}},\ }\href@noop {}
  {\bibfield  {journal} {\bibinfo  {journal} {Phys.\ Rev.\ D}\ }\textbf
  {\bibinfo {volume} {84}},\ \bibinfo {pages} {024046} (\bibinfo {year}
  {2011})},\ \Eprint {http://arxiv.org/abs/arxiv:1012.2879} {arxiv:1012.2879}
  \BibitemShut {NoStop}%
\bibitem [{\citenamefont {Zlochower}\ \emph {et~al.}(2011)\citenamefont
  {Zlochower}, \citenamefont {Campanelli},\ and\ \citenamefont
  {Lousto}}]{Zlochower:2010sn}%
  \BibitemOpen
  \bibfield  {author} {\bibinfo {author} {\bibfnamefont {Y.}~\bibnamefont
  {Zlochower}}, \bibinfo {author} {\bibfnamefont {M.}~\bibnamefont
  {Campanelli}}, \ and\ \bibinfo {author} {\bibfnamefont {C.~O.}\ \bibnamefont
  {Lousto}},\ }\href {\doibase 10.1088/0264-9381/28/11/114015} {\bibfield
  {journal} {\bibinfo  {journal} {Class.Quant.Grav.}\ }\textbf {\bibinfo
  {volume} {28}},\ \bibinfo {pages} {114015} (\bibinfo {year} {2011})},\
  \Eprint {http://arxiv.org/abs/1011.2210} {arXiv:1011.2210 [gr-qc]}
  \BibitemShut {NoStop}%
\bibitem [{\citenamefont {Lousto}\ and\ \citenamefont
  {Zlochower}(2011)}]{Lousto:2011kp}%
  \BibitemOpen
  \bibfield  {author} {\bibinfo {author} {\bibfnamefont {C.~O.}\ \bibnamefont
  {Lousto}}\ and\ \bibinfo {author} {\bibfnamefont {Y.}~\bibnamefont
  {Zlochower}},\ }\href {\doibase 10.1103/PhysRevLett.107.231102} {\bibfield
  {journal} {\bibinfo  {journal} {Phys.Rev.Lett.}\ }\textbf {\bibinfo {volume}
  {107}},\ \bibinfo {pages} {231102} (\bibinfo {year} {2011})},\ \Eprint
  {http://arxiv.org/abs/1108.2009} {arXiv:1108.2009 [gr-qc]} \BibitemShut
  {NoStop}%
\bibitem [{\citenamefont {Lousto}\ and\ \citenamefont
  {Zlochower}(2012)}]{Lousto:2012gt}%
  \BibitemOpen
  \bibfield  {author} {\bibinfo {author} {\bibfnamefont {C.~O.}\ \bibnamefont
  {Lousto}}\ and\ \bibinfo {author} {\bibfnamefont {Y.}~\bibnamefont
  {Zlochower}},\ }\href@noop {} {\  (\bibinfo {year} {2012})},\ \Eprint
  {http://arxiv.org/abs/1211.7099} {arXiv:1211.7099 [gr-qc]} \BibitemShut
  {NoStop}%
\bibitem [{\citenamefont {{Damour}}\ \emph {et~al.}(2011)\citenamefont
  {{Damour}}, \citenamefont {{Nagar}},\ and\ \citenamefont
  {{Trias}}}]{Damour:2010}%
  \BibitemOpen
  \bibfield  {author} {\bibinfo {author} {\bibfnamefont {T.}~\bibnamefont
  {{Damour}}}, \bibinfo {author} {\bibfnamefont {A.}~\bibnamefont {{Nagar}}}, \
  and\ \bibinfo {author} {\bibfnamefont {M.}~\bibnamefont {{Trias}}},\ }\href
  {\doibase 10.1103/PhysRevD.83.024006} {\bibfield  {journal} {\bibinfo
  {journal} {Phys. Rev. D}\ }\textbf {\bibinfo {volume} {83}},\ \bibinfo
  {pages} {024006} (\bibinfo {year} {2011})},\ \Eprint
  {http://arxiv.org/abs/1009.5998} {arXiv:1009.5998 [gr-qc]} \BibitemShut
  {NoStop}%
\bibitem [{\citenamefont {Santamar\'{i}a}\ \emph {et~al.}(2010)\citenamefont
  {Santamar\'{i}a}, \citenamefont {Ohme}, \citenamefont {Ajith}, \citenamefont
  {Br{\"u}gmann}, \citenamefont {Dorband}, \citenamefont {Hannam},
  \citenamefont {Husa}, \citenamefont {M{\"o}sta}, \citenamefont {Pollney},
  \citenamefont {Reisswig}, \citenamefont {Robinson}, \citenamefont {Seiler},\
  and\ \citenamefont {Krishnan}}]{Santamaria:2010yb}%
  \BibitemOpen
  \bibfield  {author} {\bibinfo {author} {\bibfnamefont {L.}~\bibnamefont
  {Santamar\'{i}a}}, \bibinfo {author} {\bibfnamefont {F.}~\bibnamefont
  {Ohme}}, \bibinfo {author} {\bibfnamefont {P.}~\bibnamefont {Ajith}},
  \bibinfo {author} {\bibfnamefont {B.}~\bibnamefont {Br{\"u}gmann}}, \bibinfo
  {author} {\bibfnamefont {N.}~\bibnamefont {Dorband}}, \bibinfo {author}
  {\bibfnamefont {M.}~\bibnamefont {Hannam}}, \bibinfo {author} {\bibfnamefont
  {S.}~\bibnamefont {Husa}}, \bibinfo {author} {\bibfnamefont {P.}~\bibnamefont
  {M{\"o}sta}}, \bibinfo {author} {\bibfnamefont {D.}~\bibnamefont {Pollney}},
  \bibinfo {author} {\bibfnamefont {C.}~\bibnamefont {Reisswig}}, \bibinfo
  {author} {\bibfnamefont {E.~L.}\ \bibnamefont {Robinson}}, \bibinfo {author}
  {\bibfnamefont {J.}~\bibnamefont {Seiler}}, \ and\ \bibinfo {author}
  {\bibfnamefont {B.}~\bibnamefont {Krishnan}},\ }\href {\doibase
  10.1103/PhysRevD.82.064016} {\bibfield  {journal} {\bibinfo  {journal}
  {Phys.\ Rev.\ D}\ }\textbf {\bibinfo {volume} {82}},\ \bibinfo {pages}
  {064016} (\bibinfo {year} {2010})},\ \Eprint {http://arxiv.org/abs/1005.3306}
  {arXiv:1005.3306 [gr-qc]} \BibitemShut {NoStop}%
\bibitem [{\citenamefont {Boyle}(2011)}]{Boyle:2011dy}%
  \BibitemOpen
  \bibfield  {author} {\bibinfo {author} {\bibfnamefont {M.}~\bibnamefont
  {Boyle}},\ }\href {\doibase 10.1103/PhysRevD.84.064013} {\bibfield  {journal}
  {\bibinfo  {journal} {Phys.\ Rev.\ D}\ }\textbf {\bibinfo {volume} {84}},\
  \bibinfo {pages} {064013} (\bibinfo {year} {2011})}\BibitemShut {NoStop}%
\bibitem [{\citenamefont {Ohme}\ \emph {et~al.}(2011)\citenamefont {Ohme},
  \citenamefont {Hannam},\ and\ \citenamefont {Husa}}]{OhmeEtAl:2011}%
  \BibitemOpen
  \bibfield  {author} {\bibinfo {author} {\bibfnamefont {F.}~\bibnamefont
  {Ohme}}, \bibinfo {author} {\bibfnamefont {M.}~\bibnamefont {Hannam}}, \ and\
  \bibinfo {author} {\bibfnamefont {S.}~\bibnamefont {Husa}},\ }\href {\doibase
  10.1103/PhysRevD.84.064029} {\bibfield  {journal} {\bibinfo  {journal}
  {Phys.\ Rev.\ D}\ }\textbf {\bibinfo {volume} {84}},\ \bibinfo {pages}
  {064029} (\bibinfo {year} {2011})}\BibitemShut {NoStop}%
\bibitem [{\citenamefont {MacDonald}\ \emph {et~al.}(2012)\citenamefont
  {MacDonald}, \citenamefont {Mroue}, \citenamefont {Pfeiffer}, \citenamefont
  {Boyle}, \citenamefont {Kidder} \emph {et~al.}}]{MacDonald:2012mp}%
  \BibitemOpen
  \bibfield  {author} {\bibinfo {author} {\bibfnamefont {I.}~\bibnamefont
  {MacDonald}}, \bibinfo {author} {\bibfnamefont {A.~H.}\ \bibnamefont
  {Mroue}}, \bibinfo {author} {\bibfnamefont {H.~P.}\ \bibnamefont {Pfeiffer}},
  \bibinfo {author} {\bibfnamefont {M.}~\bibnamefont {Boyle}}, \bibinfo
  {author} {\bibfnamefont {L.~E.}\ \bibnamefont {Kidder}},  \emph {et~al.},\
  }\href@noop {} {\  (\bibinfo {year} {2012})},\ \Eprint
  {http://arxiv.org/abs/1210.3007} {arXiv:1210.3007 [gr-qc]} \BibitemShut
  {NoStop}%
\bibitem [{SpE()}]{SpECwebsite}%
  \BibitemOpen
  \href@noop {} {}\bibinfo {howpublished}
  {\url{http://www.black-holes.org/SpEC.html}}\BibitemShut {NoStop}%
\bibitem [{\citenamefont {Boyle}\ \emph {et~al.}(2007)\citenamefont {Boyle},
  \citenamefont {Brown}, \citenamefont {Kidder}, \citenamefont {Mrou{\'e}},
  \citenamefont {Pfeiffer}, \citenamefont {Scheel}, \citenamefont {Cook},\ and\
  \citenamefont {Teukolsky}}]{Boyle2007}%
  \BibitemOpen
  \bibfield  {author} {\bibinfo {author} {\bibfnamefont {M.}~\bibnamefont
  {Boyle}}, \bibinfo {author} {\bibfnamefont {D.~A.}\ \bibnamefont {Brown}},
  \bibinfo {author} {\bibfnamefont {L.~E.}\ \bibnamefont {Kidder}}, \bibinfo
  {author} {\bibfnamefont {A.~H.}\ \bibnamefont {Mrou{\'e}}}, \bibinfo {author}
  {\bibfnamefont {H.~P.}\ \bibnamefont {Pfeiffer}}, \bibinfo {author}
  {\bibfnamefont {M.~A.}\ \bibnamefont {Scheel}}, \bibinfo {author}
  {\bibfnamefont {G.~B.}\ \bibnamefont {Cook}}, \ and\ \bibinfo {author}
  {\bibfnamefont {S.~A.}\ \bibnamefont {Teukolsky}},\ }\href {\doibase
  10.1103/PhysRevD.76.124038} {\bibfield  {journal} {\bibinfo  {journal}
  {Phys.\ Rev.\ D}\ }\textbf {\bibinfo {volume} {76}},\ \bibinfo {eid} {124038}
  (\bibinfo {year} {2007})}\BibitemShut {NoStop}%
\bibitem [{\citenamefont {{M. A. Scheel, M. Boyle, T. Chu, L. E. Kidder, K. D.
  Matthews and H. P. Pfeiffer}}(2009)}]{Scheel2009}%
  \BibitemOpen
  \bibfield  {author} {\bibinfo {author} {\bibnamefont {{M. A. Scheel, M.
  Boyle, T. Chu, L. E. Kidder, K. D. Matthews and H. P. Pfeiffer}}},\
  }\href@noop {} {\bibfield  {journal} {\bibinfo  {journal} {Phys.\ Rev.\ D}\
  }\textbf {\bibinfo {volume} {79}},\ \bibinfo {pages} {024003} (\bibinfo
  {year} {2009})},\ \Eprint {http://arxiv.org/abs/arXiv:gr-qc/0810.1767}
  {arXiv:gr-qc/0810.1767} \BibitemShut {NoStop}%
\bibitem [{\citenamefont {Chu}\ \emph {et~al.}(2009)\citenamefont {Chu},
  \citenamefont {Pfeiffer},\ and\ \citenamefont {Scheel}}]{Chu2009}%
  \BibitemOpen
  \bibfield  {author} {\bibinfo {author} {\bibfnamefont {T.}~\bibnamefont
  {Chu}}, \bibinfo {author} {\bibfnamefont {H.~P.}\ \bibnamefont {Pfeiffer}}, \
  and\ \bibinfo {author} {\bibfnamefont {M.~A.}\ \bibnamefont {Scheel}},\
  }\href {\doibase 10.1103/PhysRevD.80.124051} {\bibfield  {journal} {\bibinfo
  {journal} {Phys.\ Rev.\ D}\ }\textbf {\bibinfo {volume} {80}},\ \bibinfo
  {pages} {124051} (\bibinfo {year} {2009})},\ \Eprint
  {http://arxiv.org/abs/0909.1313} {arXiv:0909.1313 [gr-qc]} \BibitemShut
  {NoStop}%
\bibitem [{\citenamefont {Buchman}\ \emph {et~al.}(2012)\citenamefont
  {Buchman}, \citenamefont {Pfeiffer}, \citenamefont {Scheel},\ and\
  \citenamefont {Szilagyi}}]{Buchman:2012dw}%
  \BibitemOpen
  \bibfield  {author} {\bibinfo {author} {\bibfnamefont {L.~T.}\ \bibnamefont
  {Buchman}}, \bibinfo {author} {\bibfnamefont {H.~P.}\ \bibnamefont
  {Pfeiffer}}, \bibinfo {author} {\bibfnamefont {M.~A.}\ \bibnamefont
  {Scheel}}, \ and\ \bibinfo {author} {\bibfnamefont {B.}~\bibnamefont
  {Szilagyi}},\ }\href@noop {} {\bibfield  {journal} {\bibinfo  {journal}
  {Phys.\ Rev.\ D}\ }\textbf {\bibinfo {volume} {86}},\ \bibinfo {pages}
  {084033} (\bibinfo {year} {2012})},\ \Eprint {http://arxiv.org/abs/1206.3015}
  {arXiv:1206.3015 [gr-qc]} \BibitemShut {NoStop}%
\bibitem [{\citenamefont {Lovelace}\ \emph {et~al.}(2012)\citenamefont
  {Lovelace}, \citenamefont {Boyle}, \citenamefont {Scheel},\ and\
  \citenamefont {Szil\'{a}gyi}}]{Lovelace:2011nu}%
  \BibitemOpen
  \bibfield  {author} {\bibinfo {author} {\bibfnamefont {G.}~\bibnamefont
  {Lovelace}}, \bibinfo {author} {\bibfnamefont {M.}~\bibnamefont {Boyle}},
  \bibinfo {author} {\bibfnamefont {M.~A.}\ \bibnamefont {Scheel}}, \ and\
  \bibinfo {author} {\bibfnamefont {B.}~\bibnamefont {Szil\'{a}gyi}},\ }\href
  {\doibase 10.1088/0264-9381/29/4/045003} {\bibfield  {journal} {\bibinfo
  {journal} {Class. Quant. Grav.}\ }\textbf {\bibinfo {volume} {29}},\ \bibinfo
  {pages} {045003} (\bibinfo {year} {2012})},\ \Eprint
  {http://arxiv.org/abs/arXiv:1110.2229} {arXiv:arXiv:1110.2229 [gr-qc]}
  \BibitemShut {NoStop}%
\bibitem [{\citenamefont {Lovelace}\ \emph {et~al.}(2011)\citenamefont
  {Lovelace}, \citenamefont {Scheel},\ and\ \citenamefont
  {Szilagyi}}]{Lovelace:2010ne}%
  \BibitemOpen
  \bibfield  {author} {\bibinfo {author} {\bibfnamefont {G.}~\bibnamefont
  {Lovelace}}, \bibinfo {author} {\bibfnamefont {M.~A.}\ \bibnamefont
  {Scheel}}, \ and\ \bibinfo {author} {\bibfnamefont {B.}~\bibnamefont
  {Szilagyi}},\ }\href {\doibase 10.1103/PhysRevD.83.024010} {\bibfield
  {journal} {\bibinfo  {journal} {Phys.\ Rev.\ D}\ }\textbf {\bibinfo {volume}
  {83}},\ \bibinfo {pages} {024010} (\bibinfo {year} {2011})},\ \Eprint
  {http://arxiv.org/abs/1010.2777} {arXiv:1010.2777 [gr-qc]} \BibitemShut
  {NoStop}%
\bibitem [{\citenamefont {Mroue}\ and\ \citenamefont
  {Pfeiffer}(2012)}]{Mroue:2012kv}%
  \BibitemOpen
  \bibfield  {author} {\bibinfo {author} {\bibfnamefont {A.~H.}\ \bibnamefont
  {Mroue}}\ and\ \bibinfo {author} {\bibfnamefont {H.~P.}\ \bibnamefont
  {Pfeiffer}},\ }\href@noop {} {\  (\bibinfo {year} {2012})},\ \Eprint
  {http://arxiv.org/abs/1210.2958} {arXiv:1210.2958 [gr-qc]} \BibitemShut
  {NoStop}%
\bibitem [{\citenamefont {Hemberger}\ \emph {et~al.}(2012)\citenamefont
  {Hemberger}, \citenamefont {Scheel}, \citenamefont {Kidder}, \citenamefont
  {Szilagyi},\ and\ \citenamefont {Teukolsky}}]{Hemberger:2012jz}%
  \BibitemOpen
  \bibfield  {author} {\bibinfo {author} {\bibfnamefont {D.~A.}\ \bibnamefont
  {Hemberger}}, \bibinfo {author} {\bibfnamefont {M.~A.}\ \bibnamefont
  {Scheel}}, \bibinfo {author} {\bibfnamefont {L.~E.}\ \bibnamefont {Kidder}},
  \bibinfo {author} {\bibfnamefont {B.}~\bibnamefont {Szilagyi}}, \ and\
  \bibinfo {author} {\bibfnamefont {S.~A.}\ \bibnamefont {Teukolsky}},\
  }\href@noop {} {\  (\bibinfo {year} {2012})},\ \Eprint
  {http://arxiv.org/abs/1211.6079} {arXiv:1211.6079 [gr-qc]} \BibitemShut
  {NoStop}%
\bibitem [{\citenamefont {Foucart}\ \emph {et~al.}(2013)\citenamefont
  {Foucart}, \citenamefont {Deaton}, \citenamefont {Duez}, \citenamefont
  {Kidder}, \citenamefont {MacDonald}, \citenamefont {Ott}, \citenamefont
  {Pfeiffer}, \citenamefont {Scheel}, \citenamefont {Szilagyi},\ and\
  \citenamefont {Teukolsky}}]{PhysRevD.87.084006}%
  \BibitemOpen
  \bibfield  {author} {\bibinfo {author} {\bibfnamefont {F.}~\bibnamefont
  {Foucart}}, \bibinfo {author} {\bibfnamefont {M.~B.}\ \bibnamefont {Deaton}},
  \bibinfo {author} {\bibfnamefont {M.~D.}\ \bibnamefont {Duez}}, \bibinfo
  {author} {\bibfnamefont {L.~E.}\ \bibnamefont {Kidder}}, \bibinfo {author}
  {\bibfnamefont {I.}~\bibnamefont {MacDonald}}, \bibinfo {author}
  {\bibfnamefont {C.~D.}\ \bibnamefont {Ott}}, \bibinfo {author} {\bibfnamefont
  {H.~P.}\ \bibnamefont {Pfeiffer}}, \bibinfo {author} {\bibfnamefont {M.~A.}\
  \bibnamefont {Scheel}}, \bibinfo {author} {\bibfnamefont {B.}~\bibnamefont
  {Szilagyi}}, \ and\ \bibinfo {author} {\bibfnamefont {S.~A.}\ \bibnamefont
  {Teukolsky}},\ }\href {\doibase 10.1103/PhysRevD.87.084006} {\bibfield
  {journal} {\bibinfo  {journal} {Phys. Rev. D}\ }\textbf {\bibinfo {volume}
  {87}},\ \bibinfo {pages} {084006} (\bibinfo {year} {2013})}\BibitemShut
  {NoStop}%
\bibitem [{\citenamefont {Mrou\'{e}}\ \emph {et~al.}(2013)\citenamefont
  {Mrou\'{e}}, \citenamefont {Scheel}, \citenamefont {Szil\'{a}gyi},
  \citenamefont {Pfeiffer}, \citenamefont {Boyle}, \citenamefont {Hemberger},
  \citenamefont {Kidder}, \citenamefont {Lovelace}, \citenamefont {Ossokine},
  \citenamefont {Taylor}, \citenamefont {l~Zengino\u{g}lu}, \citenamefont
  {Buchman}, \citenamefont {Chu}, \citenamefont {Giesler}, \citenamefont
  {Owen},\ and\ \citenamefont {Teukolsky}}]{Mroue:ManyMergersPRL}%
  \BibitemOpen
  \bibfield  {author} {\bibinfo {author} {\bibfnamefont {A.~H.}\ \bibnamefont
  {Mrou\'{e}}}, \bibinfo {author} {\bibfnamefont {M.~A.}\ \bibnamefont
  {Scheel}}, \bibinfo {author} {\bibfnamefont {B.}~\bibnamefont
  {Szil\'{a}gyi}}, \bibinfo {author} {\bibfnamefont {H.~P.}\ \bibnamefont
  {Pfeiffer}}, \bibinfo {author} {\bibfnamefont {M.}~\bibnamefont {Boyle}},
  \bibinfo {author} {\bibfnamefont {D.~A.}\ \bibnamefont {Hemberger}}, \bibinfo
  {author} {\bibfnamefont {L.~E.}\ \bibnamefont {Kidder}}, \bibinfo {author}
  {\bibfnamefont {G.}~\bibnamefont {Lovelace}}, \bibinfo {author}
  {\bibfnamefont {S.}~\bibnamefont {Ossokine}}, \bibinfo {author}
  {\bibfnamefont {N.~W.}\ \bibnamefont {Taylor}}, \bibinfo {author}
  {\bibfnamefont {A.}~\bibnamefont {l~Zengino\u{g}lu}}, \bibinfo {author}
  {\bibfnamefont {L.~T.}\ \bibnamefont {Buchman}}, \bibinfo {author}
  {\bibfnamefont {T.}~\bibnamefont {Chu}}, \bibinfo {author} {\bibfnamefont
  {M.}~\bibnamefont {Giesler}}, \bibinfo {author} {\bibfnamefont
  {R.}~\bibnamefont {Owen}}, \ and\ \bibinfo {author} {\bibfnamefont {S.~A.}\
  \bibnamefont {Teukolsky}},\ }\href@noop {} {\bibfield  {journal} {\bibinfo
  {journal} {in preparation}\ } (\bibinfo {year} {2013})}\BibitemShut {NoStop}%
\bibitem [{\citenamefont {Scheel}\ \emph {et~al.}(2006)\citenamefont {Scheel},
  \citenamefont {Pfeiffer}, \citenamefont {Lindblom}, \citenamefont {Kidder},
  \citenamefont {Rinne},\ and\ \citenamefont {Teukolsky}}]{Scheel2006}%
  \BibitemOpen
  \bibfield  {author} {\bibinfo {author} {\bibfnamefont {M.~A.}\ \bibnamefont
  {Scheel}}, \bibinfo {author} {\bibfnamefont {H.~P.}\ \bibnamefont
  {Pfeiffer}}, \bibinfo {author} {\bibfnamefont {L.}~\bibnamefont {Lindblom}},
  \bibinfo {author} {\bibfnamefont {L.~E.}\ \bibnamefont {Kidder}}, \bibinfo
  {author} {\bibfnamefont {O.}~\bibnamefont {Rinne}}, \ and\ \bibinfo {author}
  {\bibfnamefont {S.~A.}\ \bibnamefont {Teukolsky}},\ }\href@noop {} {\bibfield
   {journal} {\bibinfo  {journal} {Phys.\ Rev.\ D}\ }\textbf {\bibinfo {volume}
  {74}},\ \bibinfo {pages} {104006} (\bibinfo {year} {2006})}\BibitemShut
  {NoStop}%
\bibitem [{\citenamefont {Pfeiffer}\ \emph {et~al.}(2007)\citenamefont
  {Pfeiffer}, \citenamefont {Brown}, \citenamefont {Kidder}, \citenamefont
  {Lindblom}, \citenamefont {Lovelace},\ and\ \citenamefont
  {Scheel}}]{Pfeiffer-Brown-etal:2007}%
  \BibitemOpen
  \bibfield  {author} {\bibinfo {author} {\bibfnamefont {H.~P.}\ \bibnamefont
  {Pfeiffer}}, \bibinfo {author} {\bibfnamefont {D.~A.}\ \bibnamefont {Brown}},
  \bibinfo {author} {\bibfnamefont {L.~E.}\ \bibnamefont {Kidder}}, \bibinfo
  {author} {\bibfnamefont {L.}~\bibnamefont {Lindblom}}, \bibinfo {author}
  {\bibfnamefont {G.}~\bibnamefont {Lovelace}}, \ and\ \bibinfo {author}
  {\bibfnamefont {M.~A.}\ \bibnamefont {Scheel}},\ }\href@noop {} {\bibfield
  {journal} {\bibinfo  {journal} {Class.\ Quantum Grav.}\ }\textbf {\bibinfo
  {volume} {24}},\ \bibinfo {pages} {S59} (\bibinfo {year} {2007})},\ \Eprint
  {http://arxiv.org/abs/gr-qc/0702106} {gr-qc/0702106} \BibitemShut {NoStop}%
\bibitem [{\citenamefont {Szilagyi}\ \emph {et~al.}(2009)\citenamefont
  {Szilagyi}, \citenamefont {Lindblom},\ and\ \citenamefont
  {Scheel}}]{Szilagyi:2009qz}%
  \BibitemOpen
  \bibfield  {author} {\bibinfo {author} {\bibfnamefont {B.}~\bibnamefont
  {Szilagyi}}, \bibinfo {author} {\bibfnamefont {L.}~\bibnamefont {Lindblom}},
  \ and\ \bibinfo {author} {\bibfnamefont {M.~A.}\ \bibnamefont {Scheel}},\
  }\href@noop {} {\bibfield  {journal} {\bibinfo  {journal} {Phys.\ Rev.\ D}\
  }\textbf {\bibinfo {volume} {80}},\ \bibinfo {pages} {124010} (\bibinfo
  {year} {2009})},\ \Eprint {http://arxiv.org/abs/0909.3557} {arXiv:0909.3557
  [gr-qc]} \BibitemShut {NoStop}%
\bibitem [{\citenamefont {Altmann}(2005)}]{altmann2005rotations}%
  \BibitemOpen
  \bibfield  {author} {\bibinfo {author} {\bibfnamefont {S.}~\bibnamefont
  {Altmann}},\ }\href@noop {} {\emph {\bibinfo {title} {Rotations, quaternions,
  and double groups}}}\ (\bibinfo  {publisher} {Dover Publications},\ \bibinfo
  {address} {Mineola, N.Y},\ \bibinfo {year} {2005})\BibitemShut {NoStop}%
\bibitem [{\citenamefont {Arribas}\ \emph {et~al.}(2006)\citenamefont
  {Arribas}, \citenamefont {Elipe},\ and\ \citenamefont
  {Palacios}}]{Arribas2006}%
  \BibitemOpen
  \bibfield  {author} {\bibinfo {author} {\bibfnamefont {M.}~\bibnamefont
  {Arribas}}, \bibinfo {author} {\bibfnamefont {A.}~\bibnamefont {Elipe}}, \
  and\ \bibinfo {author} {\bibfnamefont {M.}~\bibnamefont {Palacios}},\
  }\href@noop {} {\bibfield  {journal} {\bibinfo  {journal} {Celestial
  Mechanics and Dynamical Astronomy}\ }\textbf {\bibinfo {volume} {96}},\
  \bibinfo {pages} {239} (\bibinfo {year} {2006})}\BibitemShut {NoStop}%
\bibitem [{\citenamefont {Press}\ \emph {et~al.}(2007)\citenamefont {Press},
  \citenamefont {Teukolsky}, \citenamefont {Vetterling},\ and\ \citenamefont
  {Flannery}}]{Press2007}%
  \BibitemOpen
  \bibfield  {author} {\bibinfo {author} {\bibfnamefont {W.~H.}\ \bibnamefont
  {Press}}, \bibinfo {author} {\bibfnamefont {S.~A.}\ \bibnamefont
  {Teukolsky}}, \bibinfo {author} {\bibfnamefont {W.~T.}\ \bibnamefont
  {Vetterling}}, \ and\ \bibinfo {author} {\bibfnamefont {B.~P.}\ \bibnamefont
  {Flannery}},\ }\href@noop {} {\emph {\bibinfo {title} {Numerical Recipes: The
  Art of Scientific Computing (3rd Ed.)}}}\ (\bibinfo  {publisher} {Cambridge
  University Press},\ \bibinfo {address} {New York},\ \bibinfo {year}
  {2007})\BibitemShut {NoStop}%
\bibitem [{\citenamefont {Boyle}\ \emph {et~al.}(2011)\citenamefont {Boyle},
  \citenamefont {Owen},\ and\ \citenamefont {Pfeiffer}}]{Boyle:2011gg}%
  \BibitemOpen
  \bibfield  {author} {\bibinfo {author} {\bibfnamefont {M.}~\bibnamefont
  {Boyle}}, \bibinfo {author} {\bibfnamefont {R.}~\bibnamefont {Owen}}, \ and\
  \bibinfo {author} {\bibfnamefont {H.~P.}\ \bibnamefont {Pfeiffer}},\ }\href
  {\doibase 10.1103/PhysRevD.84.124011} {\bibfield  {journal} {\bibinfo
  {journal} {Phys.\ Rev.\ D}\ }\textbf {\bibinfo {volume} {84}},\ \bibinfo
  {pages} {124011} (\bibinfo {year} {2011})},\ \Eprint
  {http://arxiv.org/abs/arXiv:1110.2965} {arXiv:1110.2965 [gr-qc]} \BibitemShut
  {NoStop}%
\bibitem [{\citenamefont {Kidder}(1995)}]{kidder95}%
  \BibitemOpen
  \bibfield  {author} {\bibinfo {author} {\bibfnamefont {L.~E.}\ \bibnamefont
  {Kidder}},\ }\href {\doibase 10.1103/PhysRevD.52.821} {\bibfield  {journal}
  {\bibinfo  {journal} {Phys.\ Rev.\ D}\ }\textbf {\bibinfo {volume} {52}},\
  \bibinfo {pages} {821} (\bibinfo {year} {1995})}\BibitemShut {NoStop}%
\bibitem [{\citenamefont {Lovelace}\ \emph {et~al.}(2008)\citenamefont
  {Lovelace}, \citenamefont {Owen}, \citenamefont {Pfeiffer},\ and\
  \citenamefont {Chu}}]{Lovelace2008}%
  \BibitemOpen
  \bibfield  {author} {\bibinfo {author} {\bibfnamefont {G.}~\bibnamefont
  {Lovelace}}, \bibinfo {author} {\bibfnamefont {R.}~\bibnamefont {Owen}},
  \bibinfo {author} {\bibfnamefont {H.~P.}\ \bibnamefont {Pfeiffer}}, \ and\
  \bibinfo {author} {\bibfnamefont {T.}~\bibnamefont {Chu}},\ }\href@noop {}
  {\bibfield  {journal} {\bibinfo  {journal} {Phys.\ Rev.\ D}\ }\textbf
  {\bibinfo {volume} {78}},\ \bibinfo {pages} {084017} (\bibinfo {year}
  {2008})}\BibitemShut {NoStop}%
\bibitem [{\citenamefont {Buonanno}\ \emph {et~al.}(2011)\citenamefont
  {Buonanno}, \citenamefont {Kidder}, \citenamefont {Mrou\'{e}}, \citenamefont
  {Pfeiffer},\ and\ \citenamefont {Taracchini}}]{Buonanno:2010yk}%
  \BibitemOpen
  \bibfield  {author} {\bibinfo {author} {\bibfnamefont {A.}~\bibnamefont
  {Buonanno}}, \bibinfo {author} {\bibfnamefont {L.~E.}\ \bibnamefont
  {Kidder}}, \bibinfo {author} {\bibfnamefont {A.~H.}\ \bibnamefont
  {Mrou\'{e}}}, \bibinfo {author} {\bibfnamefont {H.~P.}\ \bibnamefont
  {Pfeiffer}}, \ and\ \bibinfo {author} {\bibfnamefont {A.}~\bibnamefont
  {Taracchini}},\ }\href {\doibase 10.1103/PhysRevD.83.104034} {\bibfield
  {journal} {\bibinfo  {journal} {Phys.Rev.}\ }\textbf {\bibinfo {volume}
  {D83}},\ \bibinfo {pages} {104034} (\bibinfo {year} {2011})},\ \Eprint
  {http://arxiv.org/abs/1012.1549} {arXiv:1012.1549 [gr-qc]} \BibitemShut
  {NoStop}%
\bibitem [{\citenamefont {Buonanno}\ \emph {et~al.}(2003)\citenamefont
  {Buonanno}, \citenamefont {Chen},\ and\ \citenamefont
  {Vallisneri}}]{Buonanno:2002fy}%
  \BibitemOpen
  \bibfield  {author} {\bibinfo {author} {\bibfnamefont {A.}~\bibnamefont
  {Buonanno}}, \bibinfo {author} {\bibfnamefont {Y.}~\bibnamefont {Chen}}, \
  and\ \bibinfo {author} {\bibfnamefont {M.}~\bibnamefont {Vallisneri}},\
  }\href {\doibase 10.1103/PhysRevD.67.104025, 10.1103/PhysRevD.74.029904}
  {\bibfield  {journal} {\bibinfo  {journal} {Phys.\ Rev.\ D}\ }\textbf
  {\bibinfo {volume} {67}},\ \bibinfo {pages} {104025} (\bibinfo {year}
  {2003})},\ \Eprint {http://arxiv.org/abs/gr-qc/0211087} {arXiv:gr-qc/0211087
  [gr-qc]} \BibitemShut {NoStop}%
\end{thebibliography}%
\end{document}